\documentclass[letterpaper]{article}
\usepackage{arxiv}

\usepackage[sort&compress,numbers]{natbib}

 \usepackage{url} 
\usepackage{multirow}
\usepackage{hhline}
\usepackage{graphicx}
\usepackage{subcaption}
\usepackage{setspace}
\doublespace
\usepackage{stfloats}
\usepackage{color}
\usepackage{amsmath, amsfonts, amssymb}
\usepackage{graphicx}
\usepackage[dvipsnames]{xcolor}
\usepackage{array, caption, subcaption}
\usepackage{booktabs}

\newtheorem{remark}{Remark}
\begin{document}

\title{Review of Contemporary Energy Harvesting Techniques and their Feasibility in  Wireless Geophones}
\author{Naveed Iqbal,  Mudassir Masood,  Ali Nasir, Khurram Karim Qureshi
}
\maketitle

	\begin{abstract}
	Energy harvesting converts ambient energy to electrical energy providing numerous opportunities to realize wireless sensors. Seismic exploration is a prime avenue to benefit from it as energy harvesting equipped geophones would relieve the burden of cables which account for the biggest chunk of exploration cost and equipment weight.  Since numerous energies are abundantly available in seismic fields, these can be harvested to power up geophones. However, due to the random and intermittent nature of the harvested energy, it is important that geophones must be equipped to tap from several energy sources for a stable operation. It may involve some initial installation cost but in the long run, it is cost-effective and beneficial as the sources for energy harvesting are available naturally. Extensive research has been carried out in recent years to harvest energies from various sources. However, there has not been a thorough investigation of utilizing these developments in the seismic context. In this survey,  a comprehensive literature review is provided on the research progress in energy harvesting methods suitable for direct adaptation in geophones. Specifically, the focus is on small form factor energy harvesting circuits and systems capable of harvesting energy from wind, sun, vibrations, temperature difference, and radio frequencies. Furthermore,  case studies are presented to assess the suitability of the studied energy harvesting methods. Finally,  a design of energy harvesting equipped geophone is also proposed.
	\end{abstract}

	\doublespacing

\section{Introduction}
For decades, oil and gas companies have been relying on cable-based network architectures for transmitting data from geophones to the on-site data collection center. For seismic surveys, cables are accountable for almost $50\%$ of the total cost and $75\%$ of the
total equipment weight \citep{Savazzi2008}. Data is usually collected by a large number of geophones distributed over a region
of more than $20$ km$^2$ \citep{Iqbal2021, Savazzi2009}. There has recently been a growing interest in deploying wireless geophone networks for seismic acquisition, especially in large-scale land surveys  \citep{Iqbala,Iqbal2020, Reddy2021}. The network proposed by aforementioned references consists of wireless geophones sending data to the data center directly or via gateways. However, these studies ignore the fact that connecting wires also supply power to geophones. Hence, removing the wires means that each geophone needs to be equipped with an external power supply. 

\begin{figure}[h!]
	\centering
	\includegraphics[width=9.2cm]{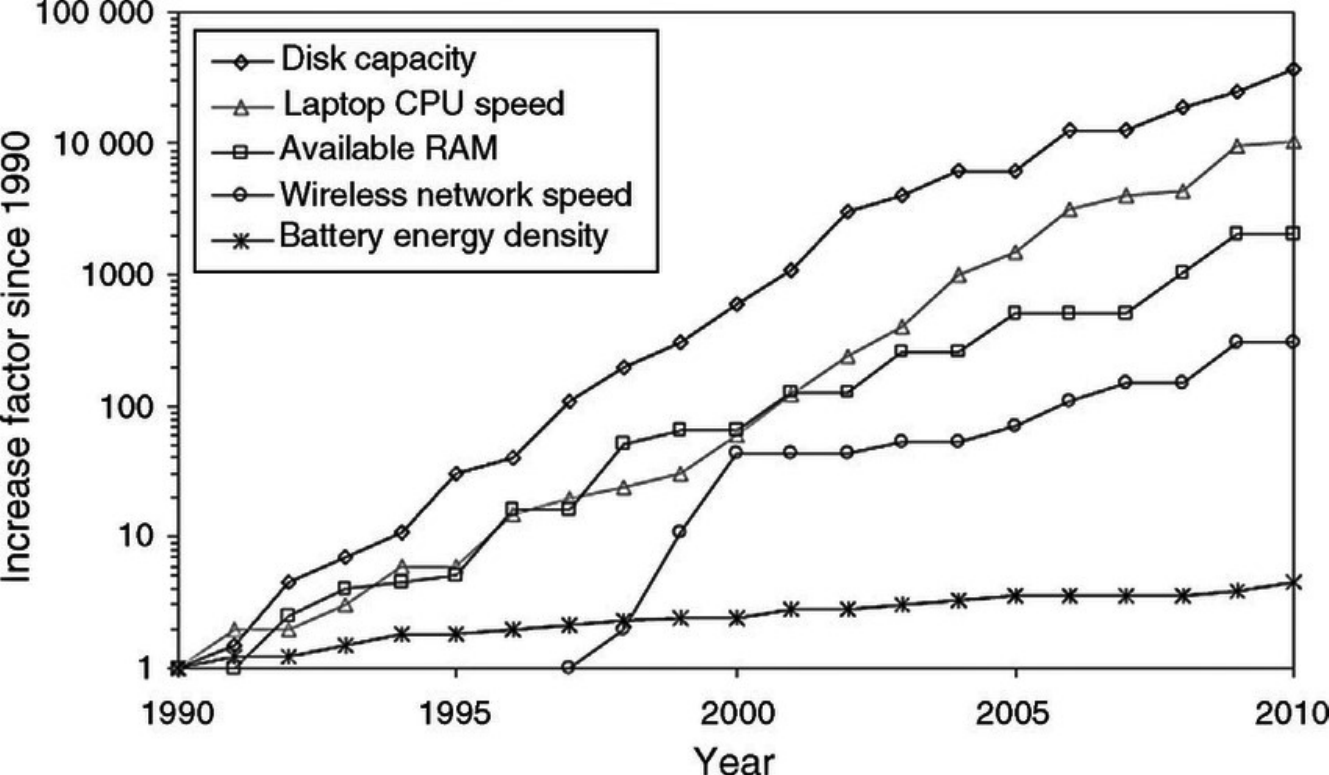}\\
	\caption{ Improvements in portable computing between 1990 and 2010. Wireless connectivity only takes into account the IEEE 802.11 standard released in 1997 (courtesy to \cite{Niroomand2016}).}
	\label{fig:imp}
\end{figure}

A commercial product available in the  market \citep{weba} has a cable-less $3$-component geophone that weighs $2.77$ lbs whereas its battery weights $2.4$ lbs. This means that $75\%$ weight has been cut-off by using the cable-free geophones, however, $86\%$ battery weight is added to the geophone weight. Hence,  batteries also account for a substantial proportion of the overall weight and size of the seismic acquisition systems negating the benefit of going cable-free. This proportion is expected to increase more as the technology scales down.  More importantly, batteries must always be recharged/replaced, and ultimately disposed of. This is a serious limitation to acquisition paradigms in which dozens or hundreds of battery-powered geophones are to be maintained. Replacement of batteries can be cumbersome and time-consuming, which may affect the seismic acquisition process.  In addition, batteries can hinder the scalability of geophone networks. The main impediments in geophone advancements are the battery's limited energy capacity and erratic lifetime efficiency. 
According to Moore’s Law, transistors double per one or two years \cite{591665}. However, the power density and life span of the batteries are limited and battery technology evolved very slowly (see Fig. \ref{fig:imp}).  Wireless geophones make it necessary to have a provision for self-powered operation. 

One of the most significant trends in electronic equipment technology since its inception has been the diminution in size and the boost in functionality. These days, small yet very powerful devices with wireless communication functionalities are commercially available. Over the past few decades, the size of the electronic circuit and
the energy required for a single (binary) operation have been dramatically reduced.  According to Moore’s law, integrated circuit technology evolves following a transistor size shrinking trend. Along with this trend, the supply voltage is also reduced due to reliability reasons. The ultimate result is a decrease in energy consumption owing to the size reduction of parasitic components. For a reduction of the scale with a factor $\alpha$ ($\alpha > 1$), the energy consumed by a particular shrunk circuit performing a specific task is decreased by
$1/\alpha^3$ \citep{GONZALEZ2002}. Advances in low-power design, therefore, open up the possibility of using energy from the environment to power electronic circuits. Therefore, to meet the energy needs of the wireless geophone systems, new sources of long-lasting and regenerative power need to be developed. 
Energy harvesting is a very appealing choice for driving the geophones, as a node's lifetime would be limited only by the failure of its own components.  

Energy harvesting is a mechanism of deriving energy from natural sources. This usually involves extracting some residual energy which could be a by-product of  an automated process or a natural phenomenon and, therefore considered as free energy \cite{electronics10060661}. Using the energy available in seismic fields would make it possible for wireless geophones to be fully self-sustaining, so that battery maintenance will eventually be eliminated. 

In this context, this work presents approaches that could be used to harvest energy in  seismic fields to power geophones. To the best of the authors' knowledge, this is the first work that addresses energy harvesting techniques with regard to geophones. The electrical energy for operating a geophone can be obtained by tapping the energy from the electromagnetic field (using radio frequency (RF)), vibrations, sunlight, wind, and temperature gradients.  These various sources of energies are abundantly available in seismic fields and can be advantageous.  Hence, the harvested energy can be used to power up a geophone directly and/or charge a small battery (or a supercapacitor, see \citep{Akbari2014}) connected to it. Various energy sources and the duration of their availability is highlighted in Table \ref{ehs}. It can be noticed here that  energies obtained using RF and temperature gradients (thermal) are available all day, so even if there is no seismic recording,  these energies are still available and can be used to recharge the geophone batteries. Wind energy depends on the speed of the wind but in general, it is also available all the time. The huge vibroseis truck (used to produce seismic waveform) generates a tremendous amount of vibration energy that can be harvested.  Vibration energy is available during the seismic shooting phases only. Therefore, the batteries may continue to charge all the time using available energy harvesting source(s), and the stored energy is then used for seismic recording and data transmission. Hence, the usual operation mode of an energy harvesting system in the seismic field implies harvesting during the peak time slots of energy availability, while the storage devices must meet the demand and supply in specified periods.

\begin{table}[]\scriptsize
	\caption{Energy harvesting sources}
	\label{ehs}
	\centering
		\begin{tabular}{cl}
		\toprule	
		 \textbf{Energy	source}           & \textbf{Availability  }                       \\ \midrule
			Solar            & During day time                     \\ 
			Vibration     &   During seismic shooting time                      \\ 
			RF           &  $24$ hours a day\\ 
			Thermal        &     $24$ hours a day                \\ 
			Wind            &    Depends on wind speed                    \\ \bottomrule
			\end{tabular}%
		\end{table}

The benefits of energy harvesting with regards to the seismic acquisition networks are multifold and include: long-lasting operability, no chemical disposal (avoids the environmental contamination), cost saving, safety, maintenance-free, no charging points, inaccessible sites operability, flexibility, scalability, ease of installation, increase lifetime, and complete removal of supply wires.  

In brief, this paper brings the following novel contributions:
\begin{itemize}

\item The detailed energy requirement analysis by the wireless geophone is provided in Section \ref{2}. The analysis incorporates all the battery-dependent tasks, e.g., sensing/recording, processing, and wireless communication. The analysis provides a baseline idea about the minimum energy to be harvested to enable continuous sensing, processing, and communication tasks.

\item Various possible energy harvesting mechanisms, that can be utilized by wireless geophones, are provided in Sections \ref{so}-\ref{rfh}. Particularly, the solar energy harvesting method with its implementation and feasibility details is discussed in Section \ref{so}. Vibration energy harvesting method with different types of such energy harvesters and their operation, design plus adequacy details are outlined in Section \ref{VEH}. Next, Section \ref{windy} is devoted to providing different means of energy harvesting through the wind. Various wind energy harvesters are discussed and their applicability  for self-powered geophones is studied. A brief survey of the thermal energy harvesting method and its workability in seismic fields is elaborated in Section \ref{ther}. Finally, Section \ref{rfh} considers the RF energy harvesting method along with highlighting its usefulness and implementation requirements for wireless geophone applications.

\item A novel design of a multi-source wireless energy harvesting geophone is proposed in Section \ref{prop}, which incorporates solar cell, antenna, piezoelectric, electromagnetic/electrostatic system, and thermoelectric generator to exploit all means of energy harvesting, i.e., solar, RF, wind, vibration, and thermal energy harvesting. It is important to mention here that the proposed design of a multi-source energy harvesting geophone can be easily modified to devise a multi-source energy harvesting based green wireless sensor networks, which can prolong the operating life of various IoT based sensor networks in the areas such as agriculture, smart cities, smart building, transportation systems, healthcare, and manufacturing. 
\end{itemize}

\section{Energy Requirement of a Geophone}\label{2}
Wireless geophones must be equipped with sensing (recording), processing, and communicating abilities. Therefore, geophones should have four key units: a sensing unit, a processing unit, a communication unit, and a power unit. The power consumed by sensing and processing units is used for data collection and data processing. Commercial geophones require an adequate amount of power to operate. For example, the geophone produced by a leading manufacturer \citep{weba} requires $115$ Wh battery for continuous recording for $30$ days ($24$ hours per day). Here the power consumption is around $159$ mW for sensing and processing. 
For computing the power consumed by the communication unit, the following approach is adopted:

The transmitted signal from the wireless geophones experience certain path-loss. Since geophones are deployed in an open-field or rural environment, the path-loss (in dB) can be modeled as follows \citep{ETSI}:
\begin{align}
\text{PL}(d) = \begin{cases}
\text{PL}_1(d), & 10 \text{m} < d < d_\text{BP} \\
\text{PL}_2(d), & d_\text{BP} < d < 10 \text{km}
\end{cases},
\end{align}
where 
$
d_{BP} = \frac{2 \pi h_\text{BS} h_u f_c}{c}
$
is the breakpoint distance, $d$ is the ground distance between the geophone and the base station (BS) either gateway or data center, $h_\text{BS}$ is the BS antenna height, $h_u$ is the height of the geophone antenna above the ground, $f_c$ is the carrier frequency, and $c$ is the speed of light,
\begin{eqnarray}
\text{PL}_1(d) &=& 20 \log_{10} \left( \frac{4 \pi d_\text{3D} f_c}{c} \right) \\
\text{PL}_2(d) &=& \text{PL}_1(d_\text{BP}) + 40 \log_{10} \left( \frac{d_\text{3D}}{d_\text{BP}} \right),
\end{eqnarray}
and $d_\text{3D} = \sqrt{ d^2 + (h_\text{BS} - h_u)^2  }$ is the 3D distance between the geophone and the BS. 

Considering the above path-loss and certain transmit power $P_t$ at the geophone, the received power $P_r$ at the BS  is given by
\begin{align}
P_r \ \text{(in dBm)} = P_t \ \text{(in dBm)}  - PL(d)
\end{align}

\begin{figure*}[t]
	\centering
	\begin{minipage}[h]{0.47\textwidth}
		\centering
		\includegraphics[width=1.01 \textwidth]{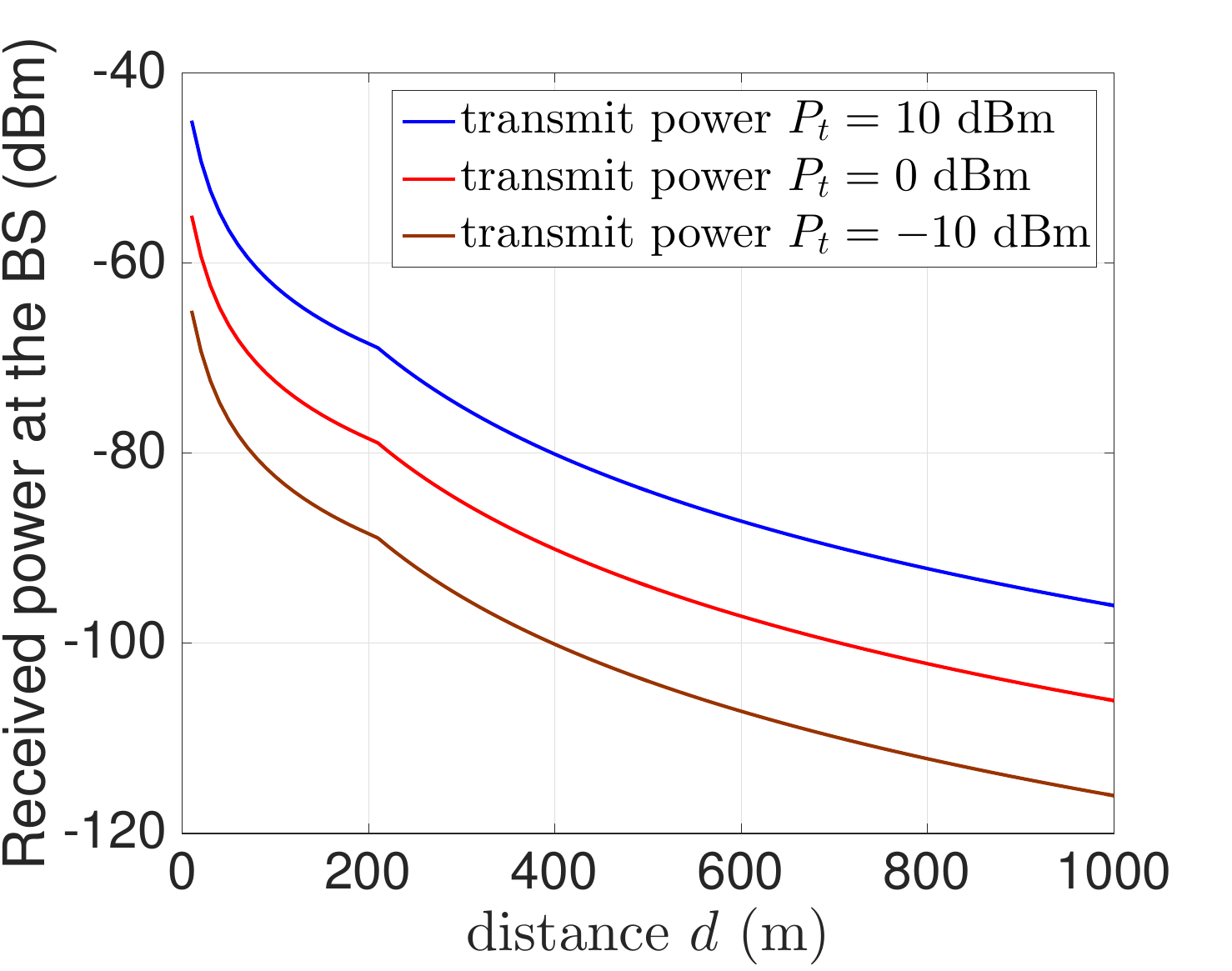}
		\caption{Received power at  BS for  $f_c = 1$ GHz, \\ $h_\text{BS} = 10$ m and $h_u = 1$ m.}\label{Rx_pow}
	\end{minipage}
	\hspace{0.3cm}
	\begin{minipage}[h]{0.47\textwidth}
		\centering
		\includegraphics[width=1.01 \textwidth]{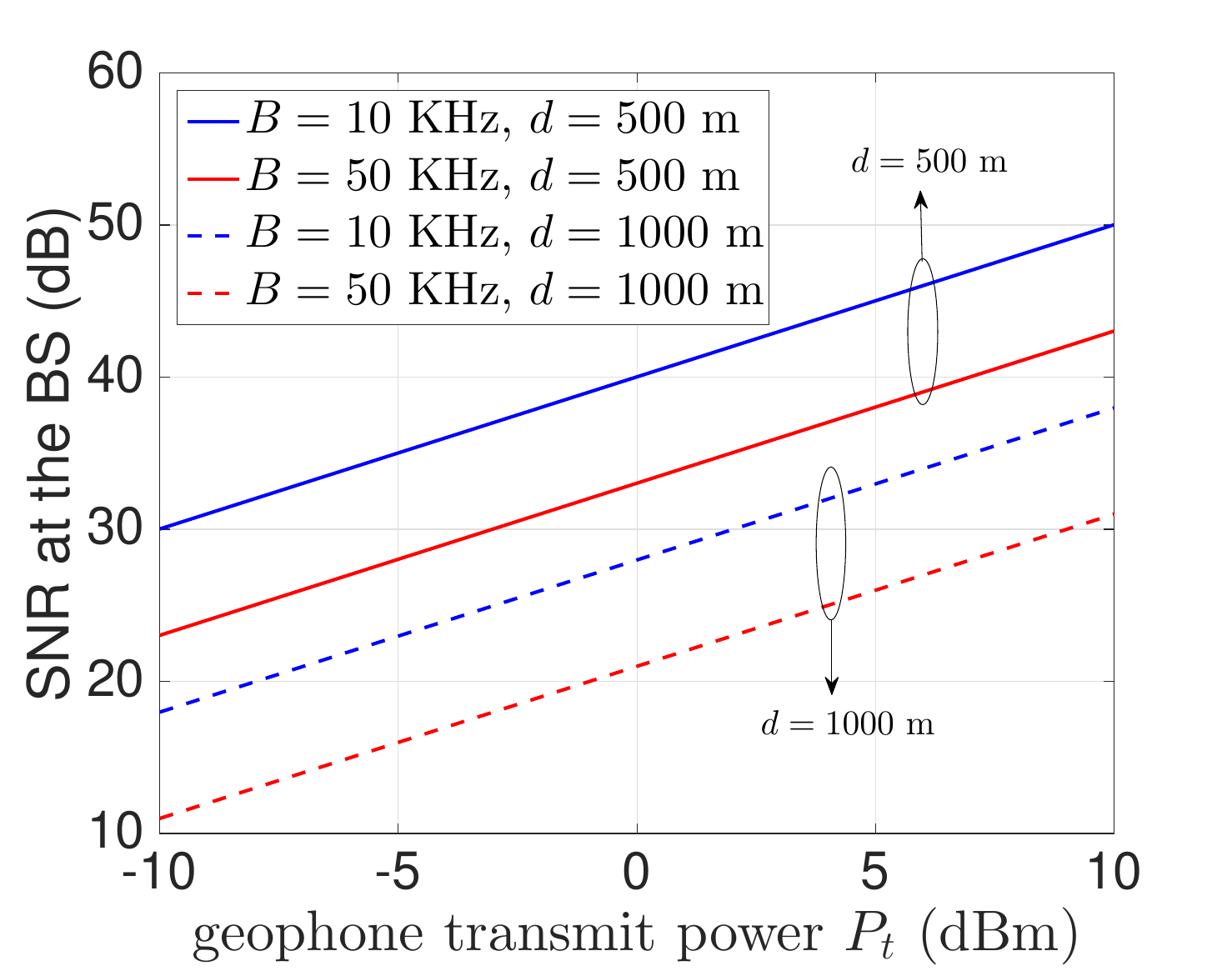}
		\caption{Received SNR at  BS for $f_c = 1$ GHz,\\  $h_\text{BS} = 10$ m and  $h_u = 1$ m.}\label{SNR}
	\end{minipage}
\end{figure*}

Considering typical values of carrier frequency $f_c = 1$ GHz, BS antenna height $h_\text{BS} = 10$ m and geophone antenna height $h_u = 1$ m, Fig. \ref{Rx_pow} depicts the BS received power $P_r$ as a function of distance $d$ for different values of the geophone transmit power $P_t$. As expected, the received power decreases with the increase in the distance $d$ due to the increase in the path-loss. Considering typical noise power desnity of $-174$ dBm/Hz, the noise power $\sigma^2$ is given by 
\[
\sigma^2  \text{(in dBm)} = -174 + 10 \log_{10}(B)
\]
where $B$ is the transmission bandwidth (BW). 

The above analysis can assist to calculate the signal-to-noise-ratio (SNR) at the BS for decoding the wireless geophone signal, which is given by
\[
\text{SNR} = P_r \ \text{(in dBm)}  - \sigma^2 \  \text{(in dBm)} .
\]

\begin{remark}
	Fig. \ref{Rx_pow} shows that if we consider the transmit power, $P_t = 0$ dBm ($1$ mW) and the ground distance of $1$ km, the received power at the BS is $-106$ dBm. This will lead to the received SNR of $28$ dB under the transmission BW of $10$ KHz (enough for achieving the data rate of $12$ kbps). This SNR is adequate to decode the signal with low bit-error rate. This implies that even assuming the quite far distance of $1$ km, we can achieve acceptable SNR of $28$ dB to decode the received signal at the BS. 
\end{remark}

Fig. \ref{SNR} plots the received SNR at the BS against different possible values of transmit power $P_r$. This figure  shows that at an extreme distance of $1$ km, the transmit power should be at least $0$ dBm to ensure SNR of $28$ dB under transmission BW of $10$ kHz.  It is clear from Fig. \ref{SNR} that the typical value of $P_t = 0$ dBm, which is equal to $1$ mW, is enough to ensure adequate SNR with sufficient coverage. 

\begin{remark}
	Now it remains how to harvest sufficient energy from different means that can allow continuous sensing and processing (power consumption of around 159 mW), and communication (transmit power requirement of around $1$ mW from the geophone to cover up to $1$ km distance). The following sections elaborate on different means that can be employed to harvest energy at the geophone.
\end{remark}

In the ensuing, various energy harvesting mechanisms are discussed and their feasibility in geophones are highlighted.
 
\section{Solar Energy Harvesting}\label{so}
The presence of a significant amount of  sunlight in outdoor environments makes it a vital energy source for geophones. A solar cell, or a photovoltaic cell, converts light energy into electricity by the photovoltaic effect. A solar cell is a tiny device made of semiconductor material. The first useful solar cell was developed in 1954 by the scientists at Bell Labs.  Since then the field of harvesting solar energy has grown steadily. A number of solar power harvesting facilities are generating hundreds of megawatts of energy exist across the world  \citep{Top19Solar}. 

Solar power is considered the most feasible form of renewable energy. This is because the Sun provides the highest energy density \citep{Mackay} as compared to other green energy sources such as wind, vibrations, etc. Outdoor solar panels can deliver energy densities in the range of 7.5 mW/cm$^2$ \citep{Mathuna2008}. Only a few wind energy harvesting methods have shown to outperform this number at very high wind speeds \citep{WindvsSolarEnergyDensity}. A number of different semiconductor materials/technologies have been used to develop solar cells. These are broadly divided into three generations as follows:
\begin{itemize}
	\item First generation: made of crystalline silicon cells
	\item Second generation: made of thin-film cells 
	\item Third generation:  made of organic, dye-sensitized, Perovskite, and multijunction cells.
\end{itemize}

The most popular of these solar/photovoltaic cells belonging to different generations and their efficiencies are listed in Table \ref{solarcelltypes}.

\begin{table}[htbp]\scriptsize
	\centering
	\caption{Characteristics of different Solar Cell Types}\label{solarcelltypes}
	\begin{tabular}{ccccc}
		\toprule
		\textbf{Generation} & \textbf{Solar Cell Type} & \textbf{Efficiency} & \textbf{Power Density} ($W/m^2$)\footnotemark & \textbf{Ref.} \\ \midrule
		
		First & Mono-crystalline & $17-18\%$ & $111-142$ & \citep{Sharma2015} \\
		
		First & Poly-crystalline & $12-14\%$ & $111-125$ & \citep{Jayakumar2009} \\
		
		Second & Amorphous Silicon & $4-8\%$ & $50-77$ &\citep{Luceno-Sanchez2019}  \\
		
		Second & Copper Indium di-Selenide & $16-23\%$ & $91-111$ & \citep{SolarFrontier2017} \\
		
		Second & Cadmium Telluride & $9-11\%$ & $77-91$ &\citep{Luceno-Sanchez2019}  \\
		
		Third & Nano-crystal/Quantum Dot & $7-9\%$ &-  & \citep{Bozyigit2015} \\
		
		Third & Polymer & $3-10\%$ &-  & \citep{Yao2019} \\
		
		Third & Dye Sensitized & $9-12\%$ &-  & \citep{Sharma2018} \\
		
		Third & Concentrated & $\approx 33-46\%$ &-  & \citep{Kinsey2013} \\
		
		Third & Perovskite & $\approx 28\%$ &-  & \citep{NREL2019,Zhao2019a} \\
		\bottomrule
	\end{tabular}
\end{table}    

\footnotetext{These values are based on the power generated by a solar module under optimal operating conditions.}

The solar cells based on the latest technology of concentrated and Perovskite material offer the highest efficiency as well as the highest energy density among all existing technologies according to lab experiments \citep{NREL2019}. However, these technologies are still in their nascence and, therefore, have several stability limitations  \citep{Matteocci2016, Juarez-Perez2018}. These technologies hold a huge potential in generating high energy but are not commercially viable. Therefore, upon their commercialization in the future, these must be considered for the application of geophones.


\subsection{Commercial Solar Cells}

Commercial solar products use either the first generation or the second generation solar cells. The solar cells belonging to the third generation are still far from being commercialized due to stability issues highlighted in the previous section. Therefore, to harvest solar power for our application of geophones, the products that are available in the market are surveyed. A number of well-known solar manufacturers currently develop solar cells having maximum efficiencies in the range of 19\% - 23\%. Some of these are listed in Table \ref{solarcellmanufacturers}  \citep{EnergySage}. 

\begin{table}[htbp]\scriptsize
	\centering
	\caption{Commercial Solar Cell Manufacturers and their Efficiencies  \citep{EnergySage}}\label{solarcellmanufacturers}
	\begin{tabular}{cccc}
		\toprule
		\textbf{Manufacturer} & \textbf{Min. Efficiency} (\%) & \textbf{Max. Efficiency} (\%) & \textbf{Avg. Efficiency} (\%) \\ \midrule
		SunPower & $16.50\%$ & $22.80\%$ & $20.70\%$ \\
		LG & $18.40\%$ & $21.70\%$ & $19.80\%$ \\
		REC Group & $15.20\%$ & $21.70\%$ & $18.11\%$ \\
		China Sunergy & $14.98\%$ & $21.17\%$ & $17.68\%$ \\
		Solaria & $19.40\%$ & $20.50\%$ & $19.76\%$ \\
		Panasonic & $19.10\%$ & $20.30\%$ & $19.65\%$ \\
		Silfab & $17.80\%$ & $20.00\%$ & $18.93\%$ \\
		Canadian Solar & $15.88\%$ & $19.91\%$ & $17.88\%$ \\
		CertainTeed Solar & $15.40\%$ & $19.90\%$ & $18.46\%$ \\
		Solartech Universal & $19.00\%$ & $19.90\%$ & $19.45\%$ \\
		JinkoSolar & $15.57\%$ & $19.88\%$ & $17.50\%$ \\
		JA Solar & $15.80\%$ & $19.80\%$ & $17.83\%$ \\
		Hanwha Q CELLS & $17.10\%$ & $19.60\%$ & $18.44\%$ \\
		Risen & $16.30\%$ & $19.60\%$ & $18.12\%$ \\
		Talesun Energy & $16.20\%$ & $19.50\%$ & $17.52\%$ \\
		\bottomrule
	\end{tabular}
\end{table}

Any solar cell used with geophones should be resilient/robust against rugged environments. Most often the geophones are exposed to extreme conditions such as high temperatures, moisture, rain, sandstorms, snow, hail, wind, etc. which may result in corrosion, significant efficiency loss, and in some cases breakdown of solar cells. Therefore, for a commercially viable solar harvesting solution,  different characteristics in addition to the solar cell efficiency need to be compared. Most notably, the following characteristics are important.

\subsubsection{Power Tolerance}

The power tolerance metric indicates the variation in the power output that could happen due to some unavoidable circumstances. These variations are measured as a percentage of the product's power rating. Most manufacturers listed in Table \ref{solarcellmanufacturers} have a 0 W negative power tolerance which means that the actual power output will always be equal to or greater than the specified output. Any product that has a non-zero negative tolerance will result in reduced power output as compared to its rating and, therefore, may not be a good choice.

\subsubsection{Temperature Coefficient}

Solar panels rely solely on the light from the Sun which is also a source of heat. Interestingly, solar panels are also sensitive to high temperatures. The output of solar panels may reduce significantly at high temperatures. The temperature coefficient indicates the rate at which the efficiency of solar panels drops for every $1^\circ$ C above $25^\circ$ C. The temperature of $25^\circ$ C is used as a reference point as all solar panel characteristics are tested at this temperature. The temperature coefficients of the solar panels from some top manufacturers are listed in Table \ref{solarcelltempcoeff}.

\begin{table}[htbp]\scriptsize
	\centering
	\caption{Temperature Coefficients of some Commercial Solar Cells  \citep{SolarTempCoeff}}\label{solarcelltempcoeff}	
	\begin{tabular}{cc}
		\toprule
		\textbf{Solar Manufacturer} & \textbf{Temperature Coefficient Range} \\ \midrule
		China Sunergy & $-0.423$ to $-0.39$\\
		Hanwha Q CELLS & $-0.42$ to $-0.37$ \\
		Hyundai & $-0.45$ to $-0.41$ \\
		LG & $-0.42$ to $-0.3$ \\
		SunPower & $-0.38$ to $-0.29$ \\
		Panasonic & $-0.3$ to $-0.29$\\
		\bottomrule
	\end{tabular}

\end{table}

\subsubsection{Durability - Snow, Hail, and Wind Load Ratings}

Our survey of several commercial products showed that when it comes to robustness against snow, hail, and wind, almost all solar panels are certified to withstand extreme conditions. We summarize the corresponding ratings in Table \ref{solarcellweatherratings}.

\begin{table}[htbp]\scriptsize
		\centering	
		\caption{Weather Ratings of Commercial Solar Cells }

	\begin{tabular}{ccc}
	\toprule	
		\textbf{Snow} & \textbf{Hail} & \textbf{Wind }\\ \midrule
		$5400$ Pa ($550$ kg/m$^2$) & $25$ mm at speed of $83$ kph &  $2400$ Pa (225 kph) \\ \bottomrule
	\end{tabular}
\label{solarcellweatherratings}
\end{table}

International Electrotechnical Commission (IEC) has proposed two standards (IEC 61215 and IEC 61646) to evaluate the reliability of solar panels. Tests are designed following the guidelines of these standards to assess the wear and tear that solar panels will experience during their lifetime. Therefore, only those panels that are certified by IEC should be selected as they are guaranteed to withstand harsh environmental conditions.

\begin{figure}
	\centering
	\includegraphics[scale=0.5]{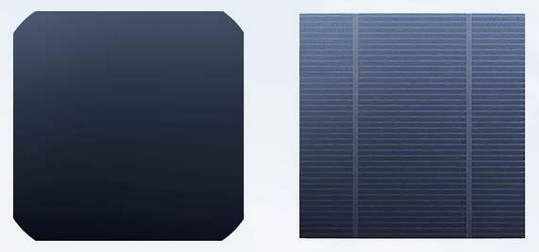}
	\caption{Front views of Maxeon (left) and a Conventional Solar Cell (right) (courtesy to \citep{Maxeon3}).}\label{fig:Maxeon}
\end{figure}

While performing the survey, we found that the solar panels based on Maxeon technology (manufactured by SunPower) stood out among all other solar panels. This is mainly due to the structural difference between Maxeon and other conventional solar cells \citep{Maxeon1, Maxeon2}. Conventional cells use busbars that run through the face of the cell to capture electrical energy created by the cell. However, Maxeon cells are backed with solid copper to capture the electrical energy as shown in Fig. \ref{fig:Maxeon}. This allows more surface area for the cell to capture energy which results in higher efficiency as evident from Table \ref{solarcellmanufacturers}. Moreover, the use of copper at the back of the cell makes it resilient to corrosion and daily wear and tear from thermal expansion, etc.

In light of the detailed review and the discussion presented above, we conclude that the solar cells based on Maxeon technology are highly efficient and at the same time robust to the harmful effects of the environment. Therefore, we propose to equip geophones with solar cells based on Maxeon technology. 

\subsection{Photodiodes}

A photodiode is also made of semiconductor material that converts light into an electric current. A photodiode is smaller than a solar cell. Its output is much lower as compared to a solar cell and is therefore mainly used as a sensor to detect light. Photodiodes have also been used to power up small electronics \citep{Satharasinghe2018}. These include mainly wearable medical sensors. However, it is not used for larger electronic devices as the generated power is not sufficient for such devices. It is due to these reasons, we have not considered photodiodes as possible sunlight energy harvesting mechanisms.

\subsection{Discussion}

The solar energy harvesting infrastructure is low cost, and noise-free. The sunlight is available to every geophone and, therefore, solar energy can be harvested by any geophone, anywhere in the world. Despite these advantages, there are some limitations. For example, sunlight is not available at night. Similarly, different weather conditions may result in limited availability of energy. Furthermore, since geophones are placed on the ground, there is a risk that solar panels would be covered by dust, and hence lowering the efficiency. Therefore, a reliable green system must not rely solely on solar energy. This implies that any reliable green solution must be hybrid, i.e., it is designed to  harness different forms of energies that are available throughout the year.

\begin{figure*}[t]
	\centering
	\begin{minipage}[h]{0.47\textwidth}
		\centering
		\includegraphics[width=1.01 \textwidth]{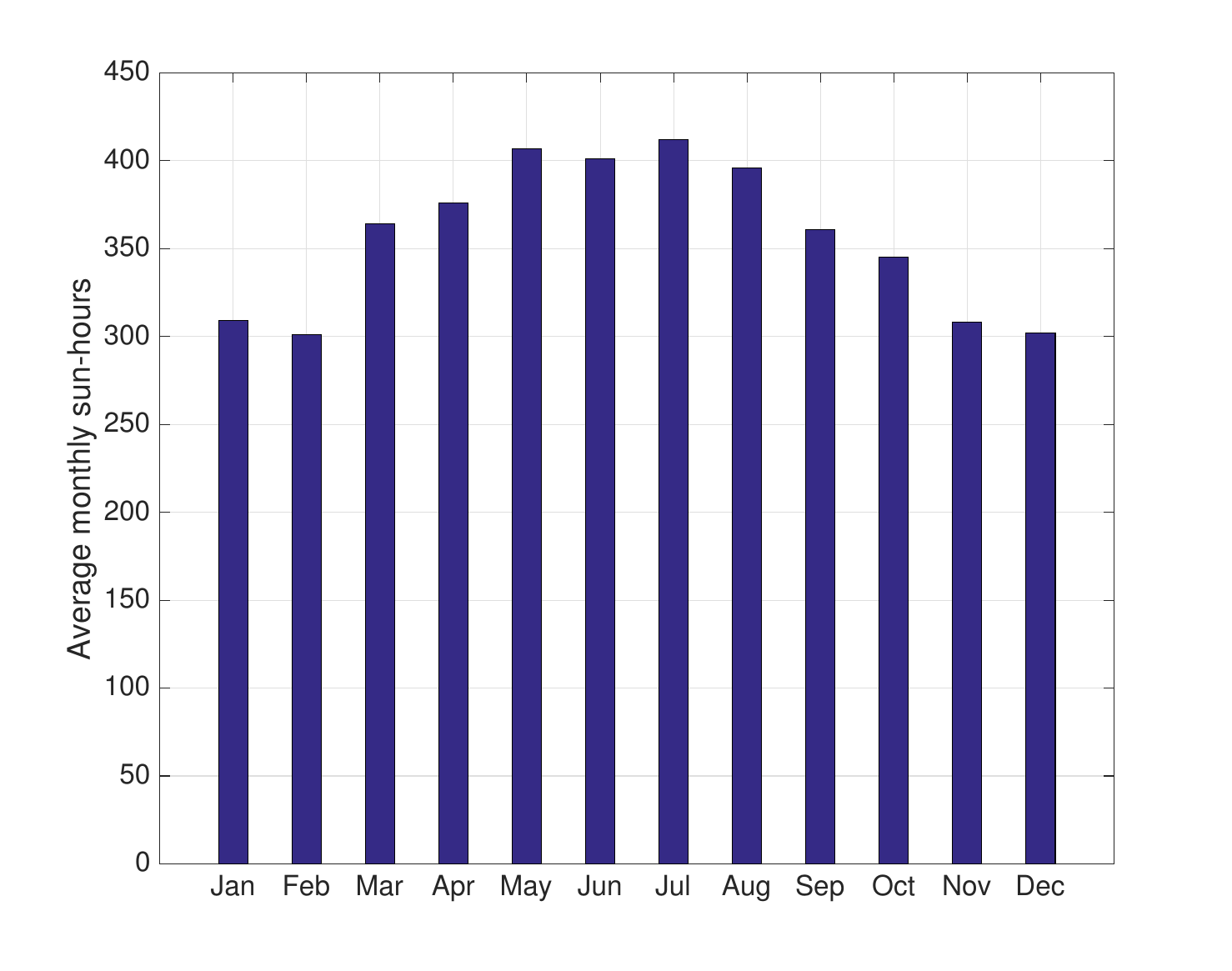}
		\caption{Average number of sun hours per month during \\ different months in Saudi Arabia, Dammam \citep{Sun_Cloud}.}\label{sunhours}
	\end{minipage}
	\hspace{0.3cm}
	\begin{minipage}[h]{0.47\textwidth}
		\centering
		\includegraphics[width=1.01 \textwidth]{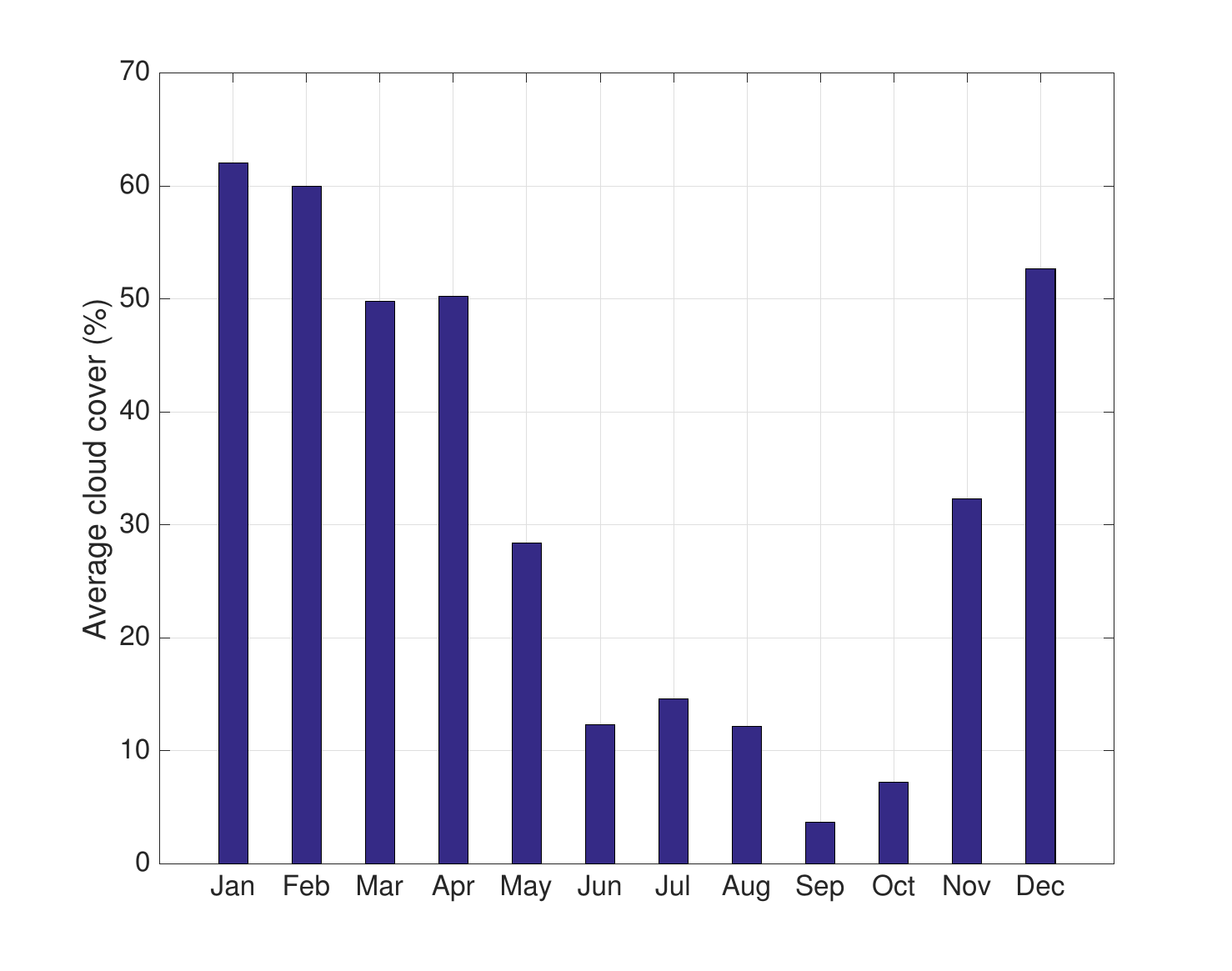}
		\caption{Average cloud coverage (in percentage) over \\ different months in Saudi Arabia, Dammam  \citep{Sun_Cloud}.}\label{cloudcover}
	\end{minipage}
\end{figure*}

As a case study, we demonstrate the viability of energy harvesting by solar energy in one of the major city  (Dammam) in the Eastern region of Saudi Arabia. Note that Saudi Arabia is chosen for the feasibility study of solar-powered wireless geophones as it is currently the largest oil producer and thus the largest consumer of geophones. The amount of harvested energy depends on the availability of the sunlight and the sky condition (whether it is clear or covered by the clouds). In this regard, Fig. \ref{sunhours} plots the average number of sun hours per month and Fig. \ref{cloudcover} plots the average cloud coverage (in percentage) during different months in Dammam, Saudi Arabia  \citep{Sun_Cloud}. It can be observed from Fig. \ref{sunhours} that sun is easily available around 12 hours per day. Fig. \ref{cloudcover} shows that the cloud coverage is also in an acceptable range. Particularly, the cloud coverage is around $10\%$ or even less during summer (June-October), which shows that solar energy harvesting is very much suitable during summer days. However, the weather is hot for most part of the year and  the temperature can reach up to $50$ $^{\circ}$C in Summer, which reduces the output of solar panels.

It is, therefore, concluded that the presence of sunlight across the world and the availability of high energy density solar cells make it feasible to equip geophones with solar cells. These solar cells could be placed around the geophone body. The surface area of geophones exposed to sunlight might be small; however, the high energy density of the cells means a sizeable amount of energy could be harvested for the successful realization of wireless geophones. Furthermore, weather is often windy in the Eastern region of Saudi Arabia and the maximum speed is $\approx 15$ m/s ($54$ km/h). It can be seen in Table \ref{solarcellweatherratings} that solar panels can withstand these harsh environmental conditions.

\section{Vibration Energy Harvesting}\label{VEH}
\begin{figure}[h!]
	\centering
	\includegraphics[width=9.2cm]{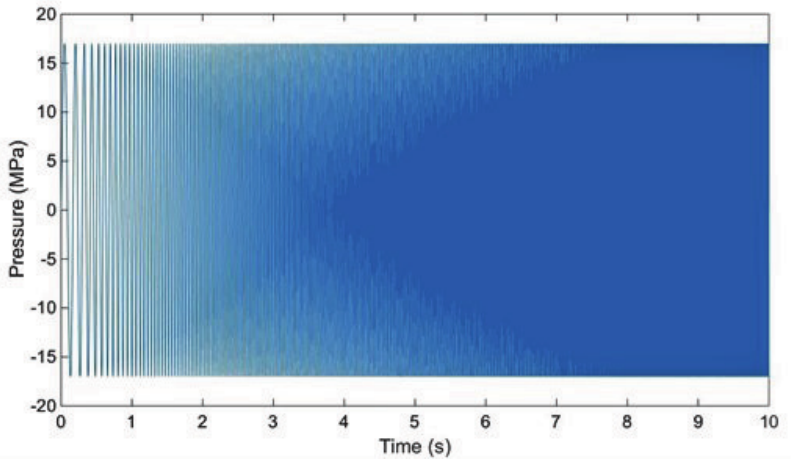}\\
	\caption{A linear sweep of $5-120$ Hz and length of $10$ sec, pressure amplitude is $17$ MPa (courtesy to \citep{Huang2017}).}
	\label{fig:Sweep}
\end{figure}

Vibration energy harvesting has been a subject of interest over the last decade. Vibration energy can be transformed into electric energy through various mechanisms, e.g., electromagnetic induction, electrostatic mechanism, or piezoelectric approach. Wireless geophones can harvest the tremendous amount of vibration energy that is generated by huge vibroseis trucks. These trucks generate vibration energy at regular intervals and thus provide a reliable source of energy to geophone. Vibroseis trucks inject a sweep (around $8$ to $10$ sec duration) of low frequencies into earth typically in the range of $1 - 100$ Hz and, therefore, it is critical to tune the energy harvesters resonant frequency accordingly.  A slight deviation could drastically reduce the amount of energy being harvested \citep{Wei2017}. An example of a linear sweep is shown in Fig. \ref{fig:Sweep}. Since the range of vibration frequency is known in our seismic scenario, the energy harvester can be designed with high efficiency.

Recently, new design approaches have been exploited based on the fact that a harvesting device's power generation performance is confined to the resonance excitation.  
In numerous applications, ambient vibration is often broadband and random, and this type of excitation must be taken into account when designing energy harvesting devices. In order words, the operating frequency bandwidth of the harvester is usually confined to a specific range that cannot cover the random vibration frequencies of external sources. Recently, researchers have  explored the  concept of broadband energy harvesting, and many  nonlinear power generators have been proposed in the literature \citep{Barton2010,Ferrari2010,Mann2010,Liu2016a,Triplett2009,Gammaitoni2009,Cottone2009,Vijayan2015}. 
 Table \ref{vib} shows a comparison of various mechanisms that are used to convert vibration energy to electrical energy. In the sequel, various vibration energy harvesting mechanisms are briefly discussed.

\subsection{Piezoelectric-based Vibration Energy Harvester}

\begin{figure}
	\centering
	\includegraphics[width=9.2cm]{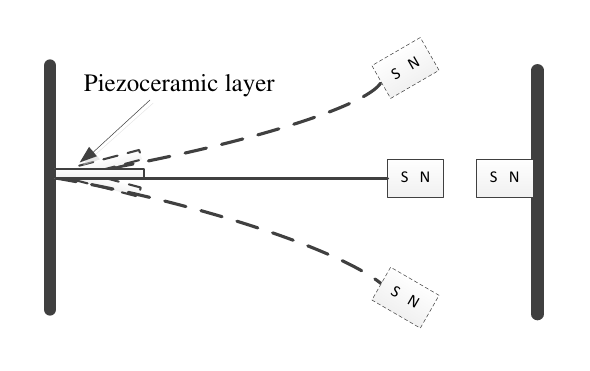}\\
	\caption{Nonlinear piezoelectric energy harvester.}
	\label{fig:piezo_mag}
\end{figure}

Piezoelectric generates an electric charge from mechanical strain. This phenomenon is known  as the direct piezoelectric effect. In the case of vibration energy harvesting, ambient vibration around the energy harvesting unit/device induces the mechanical strain.  Usually, the  piezoelectric energy harvesters of the cantilever type are developed with a proof mass located at the free end of the beam. The electric energy can be generated from bending vibrations under excitation at the root of the beam. Among various energy harvesting structures, piezoelectric transducers are one of the widely known with nonlinear characteristics, and permanent magnets are often attached to the accompanying structures to reproduce the effect of external vibration forces. 
Piezoelectric energy harvester investigated in the literature using a magnetic oscillator  \citep{Ferrari2010,Ferrari2010a,DePaula2015,Tang2012,Vocca2012,Cohen2012} is depicted in
a schematic diagram (Fig. \ref{fig:piezo_mag}).
The authors in \citep{Firoozy2017a} discovered that  harvester’s resonant frequency range is influenced by the geometric nonlinearity (in the presence or absence of the external magnets) and distance
between the magnets.   
It has been demonstrated in \citep{Zhou2017} that  the hybrid vibration energy harvesters (consisting of electromagnetic and piezoelectric generators) with nonlinear magnetic forces can effectively boost output performance  under random excitation.

\begin{table}[]
	\caption{Comparison of various mechanisms}
	\centering
	\label{vib}
	
	\resizebox{0.9\textwidth}{!}{%
	
		\begin{tabular}{llll}
			\toprule	
			& \textbf{Electrostatic }& \textbf{Electromagnetic}  & \textbf{Piezoelectric} \\ \midrule
			Material & Conductive capacitor & Neodymium iron boron  & Lead zirconate titanate \\ \hline
		 Advantages &
		 \begin{tabular}[c]{@{}l@{}}\textbullet~~Smart material not needed\\ \textbullet~~Very high output voltage ($>100$ V)\\ \textbullet~~Ease of voltage rectification and frequency tuning\end{tabular} &
		 \begin{tabular}[c]{@{}l@{}}\textbullet~~Simple construction on a large scale\\ \textbullet~~Low output impedance\\ \textbullet~~Higher output current\end{tabular} &
		 \begin{tabular}[c]{@{}l@{}}\textbullet~~Simple structure on a small scale\\ \textbullet~~High output voltage ($>5$ V)\\ \textbullet~~High coupling coefficient\end{tabular} \\ \hline 
		 Disadvantages &
		 \begin{tabular}[c]{@{}l@{}}\textbullet~~Low output current\\ \textbullet~~High impedance needed\\\textbullet~~Biased voltage required\end{tabular} &
		 \begin{tabular}[c]{@{}l@{}} \textbullet~~Small-scale limitations\\ \textbullet~~Low output voltage ($<1$ V)\\ \textbullet~~Affected by electromagnetic field\end{tabular} &
		 \begin{tabular}[c]{@{}l@{}}\textbullet~~Low output current\\ \textbullet~~Brittle\\ \textbullet~~Low strain limit\end{tabular}
		   \\ \bottomrule	    
		\end{tabular}%
	} 
\end{table}

 Piezoelectric transducers are usually manufactured using aluminum nitride, lead zirconate titanate (PZT), quartz, 
and  berlinite  \citep{Bowen2016}. New lead-free piezoelectric transducers are developed to make it more environmentally friendly, e.g., piezoelectric
nanogenerators that contain zinc nanowires (ZnO)  \citep{Qiu2012,Liao2014}.   

The piezoelectric materials are available in four types, namely thin films, single crystals, ceramics, and polymers. The piezoceramic materials,  PZT-5A, PZT-5J, and PZT-5H are used for low power applications, while for high power applications PZT-8 and PZT-4 are used  \citep{Xu2018}. Porous PZT material has the benefits of stiffness control and good capacitance.  
Piezoelectric polymers provide high power density, as do piezoelectric ceramics. However, polymer-like polyvinylidene fluoride (PVDF)  has poor adhesion to the material and a low electromechanical coupling coefficient. On the other hand, PZT is brittle and difficult to process, making it unsuitable for flexible application devices despite its high coupling coefficient  \citep{Wei2017,Jung2017}.
 
 The piezoelectric generation has received a great deal of attention due to the simple structure of the piezoelectric transducer, its compact size, and power generation efficiency. The piezo patch size is very thin, and therefore the entire
 system is simpler and smaller than  other energy harvesters  \citep{Lopes2014}. Small-scale piezoelectric transducers are also more robust and most effective
when compared to electromagnetic transducers, and hence are suitable for a compact structure such as airflow \citep{Tsujiura2017} and wind turbine energy harvester  \citep{Karami2013}, sound wave energy harvester \citep{Fang2017}, and energy harvester based on the raindrop impact \citep{Ilyas2015,Ilyas2017,Abidin2018}. In order to improve the output obtained from  piezoelectric, the authors in \citep{Abidin2018} utilized the voltage multiplier circuit in the energy harvester system. One of the issues, however,
is achieving maximum power generation efficiency  \citep{Yang2017}. Several works \citep{Yang2017a,Dhakar2013,Cook-Chennault2008}  focused on frequency bandwidth extension to maximize the efficiency. The surveyed piezoelectric energy harvesters are listed in Table \ref{PEH}.  
  
 \begin{table}[]\scriptsize
 	\caption{Piezoelectric energy harvesters}
 	\label{PEH}
 \centering
 {%
 		\begin{tabular}{c|c|c|c|c|c}
 		\toprule	
 			\textbf{Freq. (Hz)}       & \textbf{Acceleration (m/s$^2$) }& \textbf{load (k$\Omega$) }& \textbf{Max. Power output ($\mu$W)} & \textbf{AC voltage output (V)} & \textbf{Ref.} \\ \midrule
 			$126$    & $5$     & $50$            & $5.3$           & $2.6$   & \citep{Morimoto2010} \\ 
 			$113$    & $2.5$   & $200$           & $115.2$         & $8.6$   & \citep{Chen2013a} \\ 
 			$976$            & $10$              & $5.1\times10^5$  & $2.45\times10^{-5}$  & -     & \citep{Reilly2009}  \\ 
 			$1.39\times10^4$ & -               & $5.2\times10^3$  & $1$                  & $2.4$   & \citep{Jeon2005}   \\ 
 			$214$    & $19.6$  & $510$           & $1.288$         & $2.292$ & \citep{Lee2009} \\ 
 			$255.9$  & $24.5$  & $510$           & $2.675$         & $1.792$ & \citep{Lee2009} \\ 
 			$870$    & $9.8$   & -             & $1.4$           & $1.6$   & \citep{Muralt2009} \\ 
 			$572$    & $19.6$  & -             & $60$            & -     & \citep{Elfrink2009} \\ 
 			$150$    & $9.8$   & $11$            & $2.7\times10^4$ & $17$    & \citep{Wang2011a} \\ 
 			$461.15$ & $19.6$  & $6$             & $2.15$          & -     & \citep{Shen2008} \\ 
 			$125$    & $1.96$  & $1\times10^4$   & $0.12$          & -     & \citep{Tao2015} \\ 
 			$183.8$  & $7.36$  & $16$            & $0.32$          & $0.101$ & \citep{Shen2009} \\ 
 			$1700$   & -     & $5.6$           & $650$           & -     & \citep{Minazara2006} \\ 
 			$97$     & $1.96$  & $2\times10^3$   & $0.136$         & $1$     & \citep{Rezaeisaray2015} \\ 
 			$608$    & $9.8$   & $21.4$          & $2.16$          & $0.898$ & \citep{Fang2006} \\ 
 			$107$    & $2.5$   & $55.90$         & $222$           & $3.428$ & \citep{DuToit2007}  \\ 
 			$107$    & $2.5$   & $11.91$         & $586$           & $2.627$ & \citep{DuToit2007}  \\ 
 			$150$    & $5$     & $5.2\times10^3$ & $1.01$          & $2.4$   & \citep{Choi2006} \\ 
 			$2580$   & $18.36$ & $56$            & $1.8\times10^3$ & -     & \citep{Ericka2005} \\ 
 			$120$    & $2.5$   & -             & $375$           & -     & \citep{Roundy2004a}  \\ 
 			$80$     & -     & $333$           & $2$             & $1.2$   & \citep{Barton2010}  \\ 
 			$229$    & -     & -             & $3.98$          & $3.93$  & \citep{Liu2008b}  \\ \bottomrule
 		\end{tabular}%
 	}
 \end{table}

 \subsection{Electromagnetic-based Vibration Energy Harvesters}
 In an electromagnetic-based energy harvester, electrical energy is produced from the  mechanical energy  obtained by relative motion between a coil and  conductive magnetized body. The design of an electromagnetic energy harvester consists of pick-up coil, magnet, mechanical barrier arm, and cantilever beam and is used for low-frequency range applications, i.e., $1 - 10$ Hz. Downsizing of the electromagnetic harvesters with optimum power output to be adequate for low power micro-system applications is the main challenge \citep{Moss2015}.   A
 major limitation of the micro-systems is the limited number  of turns of the coil. Performance improvement is achieved by adjusting the external excitation frequency \citep{Seo2016}.
 
 The effective harvesting bandwidth can be increased by using an excitation  structure having multi-degrees of freedom system \citep{Chen2016}. Another way of making  bandwidth wider is to introduce nonlinearity in an energy harvesting system. The performance is enhanced when compared to the
 linear system. Coupling between tuning modes, hybrid transduction, and multi-modal arrays are several strategies
 used to improve efficiency through the incorporation of nonlinearity into the system \citep{Yang2017a,Tran2018}. The authors in \citep{Ooi2014} propose a
 novel electromagnetic harvester designed to improve the operating
 frequency range using the dual resonator technique. The study alluded  comprised of two separate resonator systems.
 Due to multi-vibration mode, multiple frequencies of various modes are
 tuned to a specific spectrum resulting in a wider bandwidth  \citep{Tao2014,Tao2016}.
 
Electromagnetic energy harvesters generate a good amount of
 power from weak vibration. Since power generated is  proportional to the operating frequency, the frequency-up conversion can be used in order to obtain the desired amount of average power \citep{Halim2015,Halim2015a}.
In addition, magnetoelectric transducer together with a rotary pendulum has shown to have 
frequency-doubling characteristics, hence more power is produced from low frequency   \citep{Dai2016}.
Conversely, the resonant frequency may be altered by introducing switching
damping at the expense of some power loss \citep{Ooi2015}. 
Optimal performance is observed by  Kluger \textit{et al.} \citep{Kluger2015} for small value of electromagnetic damping in the case of  linear systems and large  damping value in the case of nonlinear systems. Electromagnetic energy harvesting systems typically occupy a comparatively larger space in the devices and suffer from magnetic deterioration and windage loss. Due to the  size issue,  the fabrication of magnetic coils on micro-and nano-scales is a challenging area. Hence, the authors in  \citep{Siddique2017,Haroun2015a,Haroun2015} proposed the design based on the fact that  power increases substantially with the input amplitude, particularly with low-frequency vibrations.
 In a realistic scenario, an electromagnetic energy harvester is used to produce $30.313$ mW of power from bus vibration  \citep{Zhao2017}. 
 Electromagnetic energy harvesters perform better with larger size and periodic excitation, however, in the case of random vibration performance is weak \citep{Wang2017,Halim2016}. Various
 electromagnetic energy harvesters present in the literature are listed in Table \ref{EeEH}.
 
 \begin{table}[]\scriptsize
 	\caption{Electromagnetic energy harvesters}
 	\label{EeEH}
 \centering
{%
 		\begin{tabular}{c|c|c|c|c|c}
 			\toprule
 			\textbf{Freq. (Hz)} & \textbf{Acceleration (m/s$^2$)} & \textbf{load (k$\Omega$)} & \textbf{Max. power output ($\mu$W)} &\textbf{ AC voltage output (V)} & \textbf{Ref.} \\ \midrule
 			$322$  & $2.7$  & -              & $180$   & -                 & \citep{Glynne-Jones2004}  \\ 
 			$20.8$ & $1.96$ & $1.35\times10^3$ & $118.3$ & -                 & \citep{Bai2014} \\ 
 			$52$   & $1.7$  & -              & $120$   & -                 & \citep{Torah2008}  \\ 
 			$30$   & $1.47$ & $50$             & $20$    & $0.8$               & \citep{Wu2015} \\ 
 			$100$  & $1.96$ & -              & $240$   & -                 & \citep{Arroyo2013} \\ 
 			$12$   & $29.4$ & $40.8$           & $71.26$ & $0.47$              & \citep{Chae2013} \\ 
 			$369$  & -    & -              & $0.6$   & $1.38\times10^{-3}$ & \citep{Yang2009} \\ 
 			$62$   & -    & -              & $1.77$  & -                 & \citep{Wang2012a} \\ 
 			$30$   & -    & -              & $254$   & -                 & \citep{Wang2010} \\ 
 			$40$   & -    & -              & $153$   & -                 & \citep{Sardini2011} \\ 
 			$52$   & $0.59$ & $4$              & $46$    & -                 & \citep{Beeby2007}  \\ 
 			$128$  & -    & $6$              & $404$   & -                 & \citep{Bouendeu2011} \\ \bottomrule
 		\end{tabular}%
 	}
 \end{table}

 \subsection{Electrostatic-based Vibration Energy Harvesters} 
In electrostatic energy harvesters, charges are created by relative motion between two charged capacitor plates. This 
results in a potential difference in the capacitor and thus static electricity. Triboelectrification  refers to the transfer of the charge between two surfaces in contact. Triboelectric nanogenerators based on electrostatic induction and triboelectrification effect are invented by Fan \textit{et al.} \cite{Fan2012} in order to harvest mechanical energy from the ambient environment.

Recently, Sequeira et al.\citep{Sequeira2019} discovered the optimized capacitor plate pattern by utilizing the topological
optimization method in order to enhance the average output power.
In \citep{Park2017}, the authors designed a  single electrode mode nanogenerator for wearable products, in which silicon rubber and conductive thread are used as a negative dielectric material and  an electrode, respectively. The electrical energy is produced by interaction of the silicon layer and human skin. However, the output energy
is observed to be  very low. Improved efficiency is shown in the freestanding triboelectric setup \citep{Su2016,Zhu2012}. In  this setup, one dielectric material is  free while another pair of dielectric
material is fixed and attached to electrodes. Lateral sliding occurs between free and paired electrodes in this configuration. 
Research has been carried out in recent years to achieve optimum power output by hybridizing triboelectric materials with  electromagnetic and piezoelectric materials \citep{Askari2017,He2018}. The authors in \citep{Shao2017} have succeeded in generating energy that can supply power to LED bulbs and supercapacitors. 

Electrostatic energy harvesters require an external voltage source. The key benefit is the production of extremely high voltage due to high internal impedance as compared to other energy harvesters \citep{Aljadiri2017}. Due to the absence of smart materials like optoelectronics, piezo patches,  shape
memory alloy, and magnetostrictive, triboelectric energy harvesters are long-lasting with  adjustable coupling coefficient and low system cost. These harvesters are 
mostly used for small-scale purposes. Table \ref{EEH} depicts various
electrostatic energy harvesters present in the literature.
\begin{table}[]\scriptsize
	\caption{Electrostatic energy harvesters}
	\label{EEH}
	\centering
{%
		\begin{tabular}{c|c|c|c|c|c}
			\toprule
			\textbf{Freq. (Hz)} &\textbf{ Acceleration (m/s$^2$)} & \textbf{load (k$\Omega$)} & \textbf{Max. power output ($\mu$W)} &\textbf{ AC voltage output (V) }& \textbf{Ref.} \\ \midrule		
			$20$   & -    & $6\times10^4$  & $37.7$ & $150$    & \citep{Tsutsumino} \\ 
		    $45$  & $0.08$ & -              & $0.12$  & -    & \citep{Miyazaki}  \\ 
            $120$  & $2.25$ & -              & $116$  & -    & \citep{Roundy2002}  \\ 
            $2$    & -    & $7.0\times10^3$  & $40$   & -    & \citep{Naruse2009} \\ 
            $200$  & -    & -              & $1.6$  & -    & \citep{Torres2009} \\ 
            $96$   & $9.8$  & $1.34\times10^4$ & $0.15$ & -    & \citep{Wang2014b} \\ 
            $4.76$ & -    & $1\times10^3$    & $58$   & $24$   & \citep{TASHIRO2000} \\ 
           	$6$    & -    & -              & $36$   & -    & \citep{Tashiro2002} \\ 
            $63$   & $9.8$  & $2\times10^4 $   & $1$    & $11.2$ & \citep{Suzuki2010}\\   \bottomrule       
		\end{tabular}%
	}
\end{table}
\subsection{Discussion}
For a geophone, a hybrid vibration energy harvester can be designed. The surface area and the internal space in geophone allow us to use piezoelectric, electromagnetic, and electrostatic energy harvesters altogether. A schematic diagram of a geophone is shown in Section \ref{prop} with various types of energy harvesters. However, it should be noted that geophones close to the vibroseis truck get the maximum vibration as compared to the ones that are far away.  Hence, nearby geophones benefit more from the vibration energy harvesters for a particular shot. It is also worth mentioning that vibroseis truck moves within the seismic field  and shots are carried out at various locations to cover all the area. Roughly, we can say that every geophone gets approximately the same amount of vibration energy per day. 

Various commercial piezoelectric harvesters are available in the market to be suitable for the geophones \citep{piezo}. Among them, PPA-2011, PPA-2014, and PPA-4011 are best suited for the application at hand (see Table \ref{ppa} for PPA-4011 specifications). Furthermore, multiple piezo can be connected together for more power \citep{ppa}. 

 \begin{table}[h]\scriptsize
 	\caption{Specifications of PPA-4011 (Length $=71$ mm, Width $=25.4$ mm, Thickness $=1.3$ mm)}
 	\label{ppa}
 	\centering
 {%
 		\begin{tabular}{c|c|c|c|c|c|c}
 		\toprule	
 			\begin{tabular}[c]{@{}c@{}}\textbf{Tip Mass}\\  \textbf{(gram)}\end{tabular} &
 			\begin{tabular}[c]{@{}c@{}}\textbf{Freq.}\\  \textbf{(Hz)}\end{tabular} &
 			\begin{tabular}[c]{@{}c@{}}\textbf{Acce. Amp.} \\ \textbf{(g)}\end{tabular} &
 			\begin{tabular}[c]{@{}c@{}}\textbf{Load}\\  \textbf{(k$\mu$)}\end{tabular} &
 			\begin{tabular}[c]{@{}c@{}}\textbf{RMS Current} \\ \textbf{(mA)}\end{tabular} &
 			\begin{tabular}[c]{@{}c@{}}\textbf{RMS Voltage} \\ \textbf{(V)}\end{tabular} &
 			\begin{tabular}[c]{@{}c@{}}\textbf{RMS power}\\  \textbf{(mW)}\end{tabular} \\ \midrule
 			$ 25.3 $ & $ 63 $ & $ 0.25 $ & $ 8.1 $  & $ 0.5 $ & $ 3.9 $  & $ 1.9  $ \\ 
 			$ 25.3 $ & $ 63 $ & $ 0.50 $ & $ 8.5 $  & $ 0.8 $ & $ 6.9 $  & $ 5.6 $  \\ 
 			$ 25.3 $ &$  62 $ & $ 1.00 $ & $ 6.2 $  & $ 1.7 $ & $ 10.6 $ & $ 18.0 $ \\ 
 			$ 25.3 $ & $ 62 $ & $ 2.00 $ &$  5.0  $ & $ 3.2 $ & $ 16.2 $ & $ 52.0 $ \\ 
 			$ 28.4 $ & $ 60 $ & $ 0.25 $ & $ 7.5 $  & $ 0.5 $ & $ 4.0 $  & $ 2.1 $  \\ 
 			$ 28.4 $ & $ 60 $ &$  0.50 $ & $ 9.4 $ & $ 0.8 $ & $ 7.7 $  & $ 6.4 $  \\ 
 			$ 27.1 $ & $ 60 $ & $ 1.00 $ & $ 5.4 $ & $ 1.9 $ & $ 10.2 $ & $ 19.5  $\\ 
 			$ 26.6 $ & $ 60 $ & $ 2.00 $ & $ 4.7 $ & $ 3.5 $ & $ 16.7 $ & $ 59.0  $\\ \bottomrule
 		\end{tabular}%
 	}
 \end{table}

\section{Wind Energy Harvesting}\label{windy}
The presence of natural wind in outdoor environments makes it an important energy source for geophones. Wind energy has been used for centuries to perform different tasks. However, it is only recently that, with the advent of wireless sensor networks and the IoT, attention has been focused on miniature wind energy harvesting devices. In the past decade, the area of small-scale wind energy harvesting has gained attention and a number of designs have emerged. 

Small-scale wind energy harvesters can help achieve the goal of self-powered sensors and/or tiny devices. Therefore, these harvesters are of huge interest to realize the idea of wireless geophones. 


Wind energy can be harvested using two different mechanisms. These include:
\begin{itemize}
	\item the rotary movement of windmills/wind turbines, and 
	\item the aeroelastic behavior of materials
\end{itemize}

Most of the windmills and wind turbines work on the principle of electromagnetic induction to generate electricity. However, the rotary movement can be converted to electrical energy using other induction mechanisms as well. On the other hand, the harvesters utilizing the aeroelastic behavior of materials are mainly based on piezoelectric induction. In this section, we discuss some of the most important innovations/designs of miniature wind energy harvesters that could be successfully adopted to power geophones.

%
%
%

\subsection{Windmills and Wind Turbines}

Windmills and wind turbines are used to convert the kinetic energy of wind into mechanical energy. The mechanical energy can then be converted to electrical energy using  any of the three induction mechanisms (piezoelectric, electromagnetic or electrostatic).

Table \ref{windmillpiezoelectric} lists the major developments in the area of harvesting wind energy using wind turbines. It can be observed that all of these designs are of large dimensions (several cms) and their power density is extremely low to be useful for geophones. The design proposed in \citep{Rancourt2007} is an exception with a power density of 9.38 mW/cm$^2$. However, the efficiency of this design reduces significantly at low wind speeds. 

Actually, the efficiency of all harvesters based on rotary motion reduces drastically at lower wind speeds \citep{LowWind2019}. Designs shown in Table \ref{windmillpiezoelectric} have cut-in wind speeds in the range of 2 - 4.5 m/s with the exception of \citep{Bressers2010}. This indicates clearly that high wind speeds are needed to take advantage of the windmills and wind turbines. However, high wind speeds are not always available. Let us take an example of oil-rich Saudi Arabia where geophones find most of their usage. The average wind speed in Saudi Arabia, in general, is  6.73 m/s at a height of 100 m \citep{SaudiWindspeed2016}. However, specifically, in the oil-rich region of Saudi Arabia (Dammam), the wind speeds vary from 0.2 - 5.5 m/s \citep{earthnullschool}.  As geophones are placed on ground level, they will experience wind speeds that are much lower than the cut-in speeds.

Thus operating small-scale devices such as geophones and other sensors using such small-scale wind turbines is not a viable solution. However, other (non-rotary) designs for wind energy harvesting exist. Their merits and demerits are discussed from the point-of-view of geophones in the next sub-section.

\begin{table}[htbp]
	\caption{Summary of some prominent designs of windmills and wind turbines}\scriptsize\centering
	\begin{tabular}{ccccccc}
		\toprule
	\textbf{	Dimension (cm)} & \begin{tabular}[c]{@{}c@{}}\textbf{Power density}\\  \textbf{(mW/cm$^{3}$)}\end{tabular}  &
	\begin{tabular}[c]{@{}c@{}}\textbf{Max. power}\\  \textbf{(mW)}\end{tabular} &	 
	\begin{tabular}[c]{@{}c@{}}\textbf{Max. power speed}\\  \textbf{(m/s)}\end{tabular} & 
	\begin{tabular}[c]{@{}c@{}}\textbf{Cut-in speed}\\  \textbf{(m/s)}\end{tabular}  & \textbf{Induction Method} &\textbf{ Ref.} \\ \midrule
		
		4.2 & 9.38 & 130 & 11.8 & - & electromagnetic & \citep{Rancourt2007} \\
		
		2 & 1.368 & 4.3 & 10 & 4.5 & electromagnetic & \citep{Bansal2009}\\
		
		4.5 & 3.9 & 62.5 & 15 & - & tribolelectric & \citep{Xie2013}{$^*$} \\
		
		4 & 0.04377 & 0.55 & 20 & 4 & tribolelectric & \citep{Perez2016} \\
		
		4 & 0.0159 & 0.2 & 10 & - & electrostatic & \citep{Perez2015} \\
		
		\begin{tabular}[c]{@{}l@{}}10 bimorphs each of \\ $6 \times 2 \times 0.06$ cm$^3$\end{tabular} & 0.0663 & 7.5 & 4.5 & 2.1 & piezoelectric & \citep{Priya2005a} \\
		
		$5.08 \times 11.6 \times 7.62$ cm$^3$ & 0.0134 & 1.2 & 5.4 & 2.1 & piezoelectric & \citep{Chen2006} \\
		
		$7.62 \times 10.16 \times 12.7$ cm$^3$  & 0.0388 & 5 & 4.5 & 2.4 & piezoelectric & \citep{Myers} \\
				
		$16.51 \times 16.51 \times 22.86$ cm$^3$ & 0.00318 & 1.2 & 4.0 & 0.9 & piezoelectric & \citep{Bressers2010} \\
				
		$8 \times 8 \times 17.5$ cm$^3$ & 0.0286 & 4 & 10 & 2 & piezoelectric & \citep{Karami2013} \\ \midrule
		
	\end{tabular}\label{windmillpiezoelectric}\\
\hspace{-11cm}$^*$ open-circuit measurements
\end{table}

\subsection{Wind Energy Harvesters Utilizing Aeroelasticity}

Wind energy can be harvested by taking advantage of the aeroelastic behavior of different materials. \textit{Aeroelasticity} refers to the tendency of an elastic body to vibrate when it is exposed to a fluid flow (flow of wind/air in our case). These vibrations may be induced due to various aerodynamic phenomena such as \textit{flutter}, \textit{vortex-induced vibrations}, \textit{galloping}, and \textit{buffeting}. These phenomena are undesired  in most of the applications such as in aircraft wings, bridges, transmission lines, etc. However, these phenomena can be used to generate power. 

In this case, a wind energy harvester is exposed to a flow field which results in large limit-cycle oscillations. The kinetic energy of these oscillations may then be converted to electrical energy using either of the three transduction mechanisms. However, in almost all designs proposed in the literature, piezoelectric transduction is used due to the flexibility and efficiency that this method offers.

In the following, we list different types of harvesters that utilize various methods to take advantage of aeroelasticity. Each of these methods has been used in the literature to propose a number of harvester designs to efficiently harvest wind energy. These harvester types are:
\begin{itemize}
	\item Vortex-induced vibration (VIV) wind energy harvester
	\item Galloping energy harvester
	\item Wake Galloping energy harvester
	\item Flutter-based energy harvester
	\item Turbulence-induced vibration (TIV) wind energy harvester
\end{itemize}

The harvester designs that hold the potential to be most effective for our application of geophones are presented briefly in the following.

\subsubsection{Vortex-induced Vibration (VIV) Wind Energy Harvester}

VIV is a phenomenon in which periodic vortices are shed by a bluff body when exposed to wind. These periodic vortices cause the body to oscillate \citep{Srpkaya2004, Williamson2004}. Therefore, VIV can be used to harvest wind energy by converting the oscillations into electrical energy. Piezoelectric transduction mechanism is usually used to convert oscillations into electrical energy. The design concept is shown in Fig. \ref{fig:viv1}. Using piezoelectric material could allow VIV-based harvesters to be miniaturized without losing their capability to harvest energy at low wind speeds. Therefore, recently a microchip level energy harvester has been reported in the literature \citep{Lee2019}. The VIV energy harvesters suitable for small-scale wind energy harvesting are listed in Table \ref{cmVIV}.

\begin{table}[h!]
	\caption{VIV Energy Harvesters}\scriptsize
	\centering
	\begin{tabular}{ccccccc}
		\toprule
			\begin{tabular}[c]{@{}c@{}}\textbf{Dimensions}\\  \textbf{(Bluff - dia, len)}\end{tabular}
		 & 	 \begin{tabular}[c]{@{}c@{}}\textbf{Dimensions}\\  \textbf{(Cantilever)}\end{tabular}
		 &\begin{tabular}[c]{@{}c@{}}\textbf{Power density}\\  \textbf{(mW/cm$^{3}$)}\end{tabular}  &
		 \begin{tabular}[c]{@{}c@{}}\textbf{Max. power}\\  \textbf{(mW)}\end{tabular} &	 
		 \begin{tabular}[c]{@{}c@{}}\textbf{Max. power speed}\\  \textbf{(m/s)}\end{tabular} & 
		  \begin{tabular}[c]{@{}c@{}}\textbf{Cut-in speed}\\  \textbf{(m/s)}\end{tabular} 
		 & \textbf{Ref.}\\ \midrule
		
		2.91 cm, 3.6 cm & $3.1 \times 1.0 \times 0.0202$ cm$^3$& 1.25 $\times 10^{-3}$  & 0.03 & 5 & 3.1 & \citep{Gao2013} \\
		
		2.5 cm, 11 cm & $2.86 \times 0.63 \times 0.25$ cm$^3$& 0.0918  & 5 & 5.5 & 2 & \citep{Weinstein2012} \\
		
		1.98 cm, 20.3 cm & $26.7 \times 3.25 \times 0.0635$ cm$^3$ & 1.47 $\times 10^{-3}$  & 0.1 & 1.192 & - & \citep{Akaydin2012} \\			
			
		2 mm, 10 mm	& $2.4 \times 2.4 \times 0.01$ mm$^3$ & 5.66 $\times 10^{-6}$  & 1.6 $\times 10^{-6}$ & 4.48 & - & \citep{Lee2019} \\ \midrule
		
	\end{tabular}\label{cmVIV}
\end{table}

\subsubsection{Galloping Vibrations Wind Energy Harvester}

Galloping vibrations could be induced by replacing the smooth cylindrical bluff body in the schematic of Fig. \ref{fig:viv1} with a prismatic body as shown in Fig. \ref{fig:galloping1}. The prisms used to generate galloping vibrations could be of different shapes; for example, rectangular, triangular, D-shape, hexagonal, rectangular with a V-shape groove, etc. The galloping-induced vibrations have large amplitudes  as compared to VIV and lend themselves for the stable acquisition of energy \citep{Li2016}.  Galloping vibrations can be harvested using methods similar to those for VIV. Table \ref{Galloping} summarizes the performance of some prominent harvesters with different types and dimensions of the prismatic bluff bodies. It can be observed that galloping vibrations-based wind energy harvesters feature low cut-in speeds and are, therefore, useful in environments with slow wind.

\begin{table}[h!]
	\caption{Galloping Vibration Wind Energy Harvesters}\scriptsize\centering
	\begin{tabular}{cccccccc}
		\toprule
		\textbf{\begin{tabular}[c]{@{}l@{}}Bluff \\  Shape\end{tabular}} &
		\textbf{\begin{tabular}[c]{@{}l@{}}Dimensions \\  (Bluff -sides, len)\end{tabular}} &
		\textbf{\begin{tabular}[c]{@{}l@{}}Dimensions \\  (Cantilever -cm3)\end{tabular}} &
		\textbf{\begin{tabular}[c]{@{}l@{}}Power density\\ (mW/cm3)\end{tabular}} &
		\textbf{\begin{tabular}[c]{@{}l@{}}Max. power \\  (mW)\end{tabular}} &
		\textbf{\begin{tabular}[c]{@{}l@{}}Max. power \\  speed (m/s)\end{tabular}} &
		\textbf{\begin{tabular}[c]{@{}l@{}}Cut-in speed \\  (m/s)\end{tabular}} &
		\textbf{Ref.} \\ \midrule
		
		\begin{tabular}[c]{@{}l@{}}V-shaped \\  groove\end{tabular}  & $2.91$ cm, $3.6$ cm & $3.1 \times 1.0 \times 0.0202$ & 1.25 $\times 10^{-3}$  & $0.03$ & $5$ & $3.1$ & \citep{Zhao2019} \\
		
		Triangle & \begin{tabular}[c]{@{}l@{}}base=$50$ cm,\\ sides=$65$ cm,\\ len=$160$ cm \end{tabular} & -  & $4.464$ & $2.4$ & $15$ & $0.336$ & \citep{Tan2019} \\
		
		Triangle & $4$ cm, $25.1$ cm & $16.1 \times 3.8 \times 0.0635$ & $0.281$   & $50$ & $5.2$ & $3.6$ & \citep{Sirohi2011} \\
		
		D-shape & $3$ cm, $23.5$ cm & $9 \times 3.8 \times 0.0635$ & $0.0134$  & $1.14$ & $4.7$ & $2.5$ & \citep{Sirohi2012} \\
		
		Square & $4$ cm, $15$ cm & \begin{tabular}[c]{@{}l@{}}inner: $5.7 \times 3 \times 0.03$,\\ outer: $17.2 \times 6.6 \times 0.06$\end{tabular} & $0.0162$   & $4$ & $5$ & $1$ & \citep{Zhao2014} \\			
				
		Square & $2$ cm, $10$ cm & $13 \times 2 \times 0.06$ & $0.0782$   & $3.25$ & $7$ & $3$ & \citep{Zhao2015, Zhao2017a}\\ \bottomrule			
			
	\end{tabular}\label{Galloping}
\end{table}

\subsubsection{Wake Galloping Vibrations Wind Energy Harvester}

Among various aerodynamic phenomena, wake galloping is the most suitable for a wind energy harvesting system. Such systems have low cut-in speeds and can operate on a wide range of wind speeds \citep{Jung2011}.  Wake galloping occurs when a fixed cylinder is placed in front of another cylinder with a flexible base. Due to the wakes from the windward cylinder, the flexible base cylinder vibrates significantly. The vibrations of the downstream cylinder are called wake galloping. These vibrations are significantly higher than those occurring due to the galloping phenomenon discussed earlier. Fig. \ref{fig:wakegalloping2} shows the schematic of a wake galloping-based wind energy harvesting system. Some prominent wake galloping-based harvesting designs are summarized in Table \ref{WakeGalloping}.

\begin{table}[h!]
	\caption{Wake Galloping Vibration Wind Energy Harvesters}\scriptsize\centering
	\begin{tabular}{cccccccc}
		\toprule
		\textbf{\begin{tabular}[c]{@{}c@{}}Bluff \\ Shape \end{tabular}} & \textbf{\begin{tabular}[c]{@{}c@{}}Bluff \\ Dimensions \end{tabular}} & \textbf{Spacing} &\begin{tabular}[c]{@{}c@{}}\textbf{Power density}\\  \textbf{(mW/cm$^{3}$)}\end{tabular}  &
		\begin{tabular}[c]{@{}c@{}}\textbf{Max. power}\\  \textbf{(mW)}\end{tabular} &	 
		\begin{tabular}[c]{@{}c@{}}\textbf{Max. power}\\   \textbf{speed (m/s)}\end{tabular} & 
		\begin{tabular}[c]{@{}c@{}}\textbf{Cut-in speed}\\  \textbf{(m/s)}\end{tabular} 
		& \textbf{Ref.}\\ \midrule
		
		\begin{tabular}[c]{@{}c@{}}Both: circular\\ cylinder\end{tabular} & \begin{tabular}[c]{@{}c@{}}Both: dia = $5$ cm, \\len = $85$ cm\end{tabular} & $25$ cm & $0.111$  & $370.4$ & $4.5$ & $1.2$ & \citep{Jung2011} \\
		
		\begin{tabular}[c]{@{}c@{}}Inner: square cylinder,\\ outer: circular cylinder \end{tabular}& \begin{tabular}[c]{@{}c@{}}Inner: sides = $1.28$ cm,\\ len = $26.67$ cm,\\ Outer: dia = $1.25$ cm,\\ len = $27.15$ cm\end{tabular} & $24$ cm & $0.572 \times 10^{-3}$  & $0.05$ & $3.05$ & $0.4$ & \citep{Abdelkefi2013} \\
			
		\begin{tabular}[c]{@{}c@{}}Both: circular \\cylinder\end{tabular} & \begin{tabular}[c]{@{}c@{}}Both: dia = $0.3$ cm,\\ len = $25$ cm \end{tabular} & $15$ cm & -  & - & - & 4 & \citep{Usman2018} \\ \bottomrule
							
	\end{tabular}\label{WakeGalloping}
\end{table}

\subsubsection{Flutter-induced Vibration Wind Energy Harvester}

\begin{figure}
	\centering
	\begin{subfigure}{.5\textwidth}
		\centering
		\includegraphics[width=.85\linewidth]{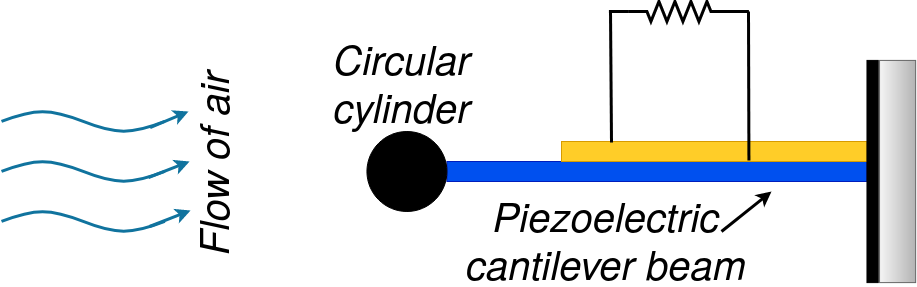}
		\caption{VIV-based harvester}
		\label{fig:viv1}
	\end{subfigure}%
	\begin{subfigure}{.5\textwidth}
		\centering
		\includegraphics[width=.85\linewidth]{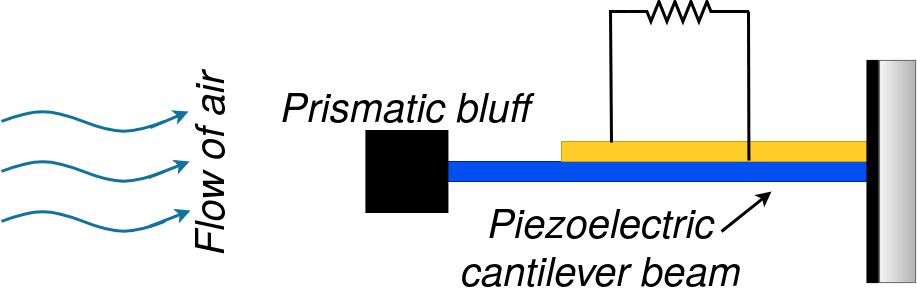}
		\caption{Galloping-based harvester}
		\label{fig:galloping1}
	\end{subfigure}\\
	\vspace{0.5cm}
	\begin{subfigure}{.5\textwidth}
		\centering
		\includegraphics[width=.85\linewidth]{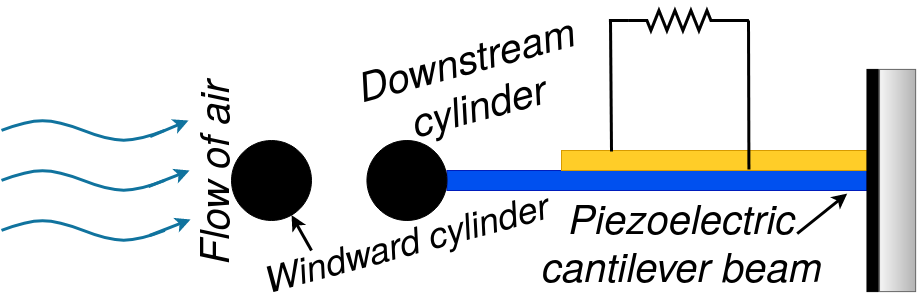}
		\caption{Wake galloping-based harvester}
		\label{fig:wakegalloping2}
	\end{subfigure}%
	\begin{subfigure}{.5\textwidth}
		\centering
		\includegraphics[width=.85\linewidth]{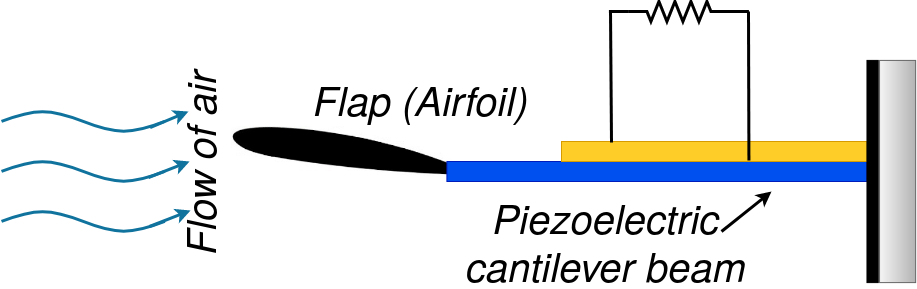}
		\caption{Flutter-based harvester}
		\label{fig:Flutter1}
	\end{subfigure}
	
	\caption{Schematic of piezoelectric energy harvester based on different aeroelasticity phenomena}
	\label{fig:test}
\end{figure}

The schematic of Fig. \ref{fig:Flutter1} shows a typical system to harness wind energy using the aerodynamic phenomenon of flutter. In this system, a flap (airfoil) is connected at the end of a cantilever. The flow of air causes the flap to flutter and the resulting vibrations can be converted to electrical energy.

The cut-in wind speeds of flutter-based systems are usually high ($> 10$ m/s). Therefore, these systems are suitable for high-wind regimes in general. However, some designs having normal cut-in wind speeds (2 - 4 m/s) exist and are summarized in Table \ref{Fluttering}.

\begin{table}[h!]
	\caption{Fluttering Vibration Wind Energy Harvesters}\scriptsize\centering
	\begin{tabular}{cccccccc}
		\toprule
		\textbf{\begin{tabular}[c]{@{}c@{}}Flap \\Type\end{tabular}} & \textbf{\begin{tabular}[c]{@{}c@{}}Flutter\\ Instability\end{tabular}} & \textbf{Dimensions} & \begin{tabular}[c]{@{}c@{}}\textbf{Power density}\\  \textbf{(mW/cm$^{3}$)}\end{tabular}  &
		\begin{tabular}[c]{@{}c@{}}\textbf{Max. power}\\  \textbf{(mW)}\end{tabular} &	 
		\begin{tabular}[c]{@{}c@{}}\textbf{Max. power}\\   \textbf{speed (m/s)}\end{tabular} & 
		\begin{tabular}[c]{@{}c@{}}\textbf{Cut-in speed}\\  \textbf{(m/s)}\end{tabular} 
		& \textbf{Ref.}\\ \midrule
		Airfoil & 	\begin{tabular}[c]{@{}c@{}}Modal\\ convergence\end{tabular} & 	\begin{tabular}[c]{@{}c@{}} Airfoil: semichord $2.97$ cm,\\ span $13.6$ cm. Cantilever:\\ $25.4 \times 2.54 \times 0.0381$ cm$^3$\end{tabular} & $7.17$ $\times 10^{-3}$  & $2.2$ & $7.9$ & $1.86$ & \citep{Bryant2011} \\
		
		Flat plate & 	\begin{tabular}[c]{@{}c@{}}Modal\\ convergence\end{tabular} & 	\begin{tabular}[c]{@{}c@{}}Flat plate: $6 \times 3$ cm$^2$.\\ Cantilever: $10 \times 6 \times 0.02$ cm$^3$\end{tabular} & $2.56$ & $4$ & $15$ & $4$ & \citep{Kwon2010} \\
				
		Flat plate & 	\begin{tabular}[c]{@{}c@{}}Modal\\ convergence\end{tabular} & 	\begin{tabular}[c]{@{}c@{}}Flat plate: $3 \times 2$ cm$^2$.\\ Cantilever: $4.2 \times 3 \times 0.01016$ cm$^3$\end{tabular} & $5.82$ & $1.1$ & $8$ & $4$ & \citep{Park2014} \\ \bottomrule					
	\end{tabular}\label{Fluttering}
\end{table}

\subsubsection{Turbulence-induced Vibration (TIV) Wind Energy Harvester}

Turbulence-induced vibrations-based wind energy harvesters are one of the most practical small-scale solutions. Almost all of the above-mentioned designs work in laminar flow conditions. In other words, if the airflow is turbulent then these harvesters will become extremely inefficient and most often seize to harvest energy. Therefore, it is natural to design harvesters that can still work in turbulent winds. Another limitation of the harvesters mentioned earlier is that they generate vibrations (and generate electricity) only if the wind speed is above a minimum limit (cut-in speed). However, interestingly, turbulence-induced vibrations (TIVs) occur even if the wind speed is very low. This phenomenon can be used to design efficient turbulence-induced vibration wind energy harvesters \citep{Hobeck2012, Hobeck2014, Akaydin2010, Liu2014, Liu2012a, He2013a}. Some of the prominent efforts in this direction are summarized in Table \ref{mmTIV}.

\begin{table}[htbp]
	\caption{mm-Scale TIV Energy Harvesters}\scriptsize\centering
	\begin{tabular}{cccccc}
		\toprule
		\textbf{Dimensions}& \begin{tabular}[c]{@{}c@{}}\textbf{Power density}\\  \textbf{(mW/cm$^{3}$)}\end{tabular}  &
		\begin{tabular}[c]{@{}c@{}}\textbf{Max. power}\\  \textbf{($\mu$W)}\end{tabular} &	 
		\begin{tabular}[c]{@{}c@{}}\textbf{Max. power}\\   \textbf{speed (m/s)}\end{tabular} & 
		\begin{tabular}[c]{@{}c@{}}\textbf{Cut-in speed}\\  \textbf{(m/s)}\end{tabular} 
		& \textbf{Ref.}\\ \midrule
		
		\begin{tabular}[c]{@{}c@{}}Bluff: $4.45 \times 4.45 \times 10.92$ cm$^3$,
		Cantilever:\\ $0.1016 \times 0.0254 \times (0.1016\times 10^{-3})$ cm$^3$\end{tabular} & 0.0184 & 4 $\times 10^{3}$ & $11.5$ & $9$ & \citep{Hobeck2012} \\
						
		\begin{tabular}[c]{@{}c@{}}Bluff: dia=$3$ cm, len=$1.2$ m,
		Cantilever:\\ $3 \times 1.6 \times 0.02$ cm$^3$\end{tabular} & $6.48 \times 10^{-8}$ & $0.055$ & $11$ & $5$ &  \citep{Akaydin2010} \\
			
		whole body: $2 \times 3.3 \times 0.4$ mm$^3$ & 28.5 $\times 10^{-3}$ & 38.7 $\times 10^{-3}$ & $15.6$ & $3.2$ & \citep{Liu2014} \\
				
		PZT beam: $3 \times 0.3 \times 0.008$ mm$^3$ & $0.36$ & $3.3$ $\times 10^{-3}$ & $15.6$ & - & \citep{Liu2012a} \\
						
		\begin{tabular}[c]{@{}c@{}}Bluff: $3 \times 7 \times 0.51$ mm$^3$
		Cantilever:\\ $3 \times 8 \times 0.035$ mm$^3$\end{tabular} & $6.3$ & $2.27$  & $16.3$  & - &  \citep{He2013a} \\ \bottomrule
				
	\end{tabular}\label{mmTIV}
\end{table}

It can be seen from the data provided in Table \ref{mmTIV} that the designs in \citep{Liu2014, Liu2012a, He2013a} are most suitable for tiny devices. Moreover, the power density of these designs is also high when compared to other designs.

\subsection{Discussion}

It is important to note that, during this survey, we could not find any small/tiny-scale commercial solution for wind energy harvesting. However, the survey gives an interesting glimpse into the flurry of activity happening towards achieving tiny-scale wind energy harvesting solutions. These solutions are mainly targeted towards IoT sensors. We believe that the application of seismic energy harvesting provides a great incentive to the industry to seriously look into commercializing some of these wind energy harvesting ideas.

\begin{figure}[htbp]
	\centering
	\includegraphics[width=\textwidth]{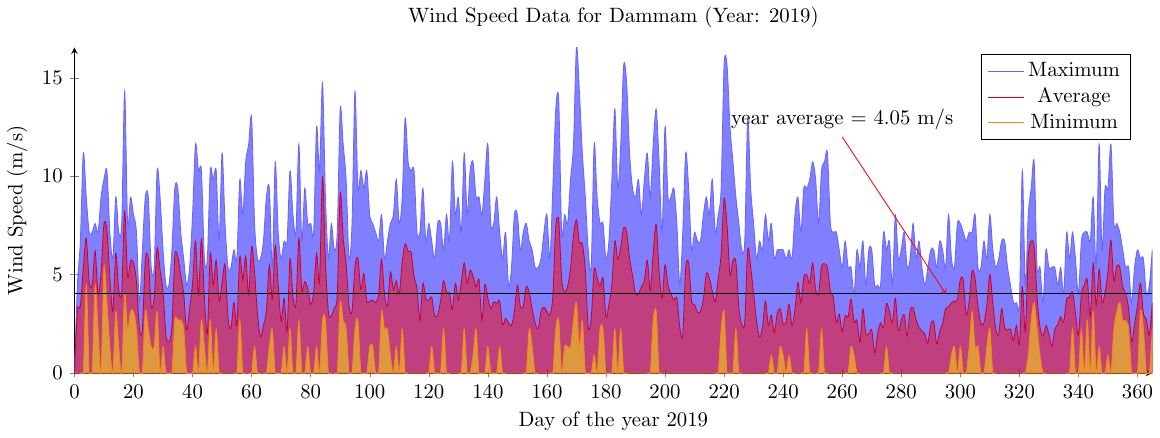}
	\caption{Wind speed data for Dammam city for  the year 2019}
	\label{Dammam2019}
\end{figure}

\begin{figure}[htbp]
	\centering
	\includegraphics[width=10cm]{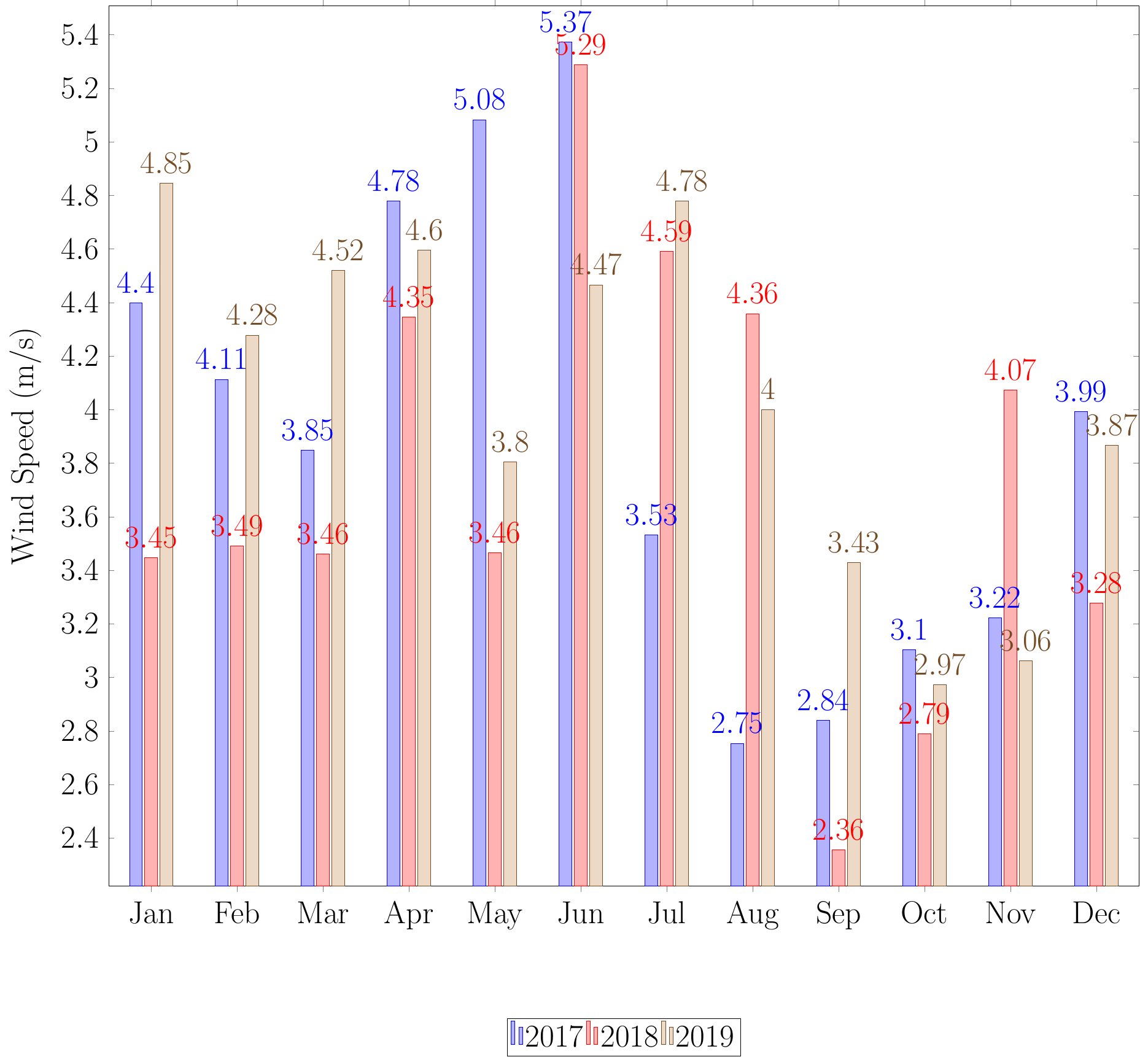}
	\caption{Average wind speeds in Dammam city for three-year period (2017-2019)}
	\label{ThreeYearBarGraph}
\end{figure}

Further, the results of the survey show that most of the wind energy harvesting methods do not perform efficiently at low wind speeds. Thus such techniques are not suitable for regions with low average wind speeds. As an example, the wind speed data of Dammam city in Saudi Arabia is presented. Figure \ref{Dammam2019} shows the maximum, minimum, and average wind speeds in Dammam for each day of the year 2019. It could be easily noted that while maximum wind speeds are as high as 15 m/s, the average speed every day is around 4 m/s. Similarly, Fig. \ref{ThreeYearBarGraph} provides a snapshot of the average wind speeds over the period of three years (2017-2019) again for Dammam city. With this data, it is obvious that for a wind energy harvesting system to be effective for Dammam city, the cut-in wind speed must be less than 4 m/s. All wind data has been acquired from the Weather Underground website \citep{wunderground}.

Moreover, as the amount of energy generated by green energy harvesting solutions is not sufficient for the sustainable operation of a geophone, it is important to devise a hybrid system. Therefore, wind energy harvesting could be used along with other energy harvesting methods discussed in this paper to provide a sustainable solution.

\section{Thermal Energy Harvesting}	\label{ther}
Another possible solution to power geophones is through energy harvesting from the temperature gradient that exists between the part of a geophone inserted inside the ground and the part that is exposed to the open environment in the seismic field. Thermal energy harvesting is coined as the reliable conversion of thermal energy to electricity with no moving parts. There are various strategies of thermal energy harvesting reported in the literature \citep{XinLu2010,Chen2016a,therm,DW,Franciscatto2014,Wu2012,Pearson2012,Sigrist2020}. However, the most notable are pyroelectric and thermoelectric generators. The first type, called pyroelectric generator, converts the temperature fluctuations in material to usable electrical energy. On the other hand, the thermoelectric generators, do not require temperature fluctuations; rather they rely on the temperature differences. In this mode of energy generation, the thermal gradient is converted into useful electrical energy utilizing the phenomenon termed as Seebeck effect which is summarized as follows. 

When two dissimilar electrical conductors are joined together, a thermocouple is formed. An electromotive force is developed when the temperature difference is maintained between the two joining junctions. The induced voltage is proportional to the temperature gradient. The heat source provides an elevated temperature where the heat flows through a thermoelectric converter to a heat sink, which is maintained at a temperature well below that of the source. Hence, the flow of charge carriers between the hot and cold bodies creates voltage difference, leading to power generation. 

The thermoelectric generators offer unique characteristics, such as; small footprint, lightweight, solid-state with no moving parts, free from noise, resistant to mechanical damage which means less maintenance, and long-term use in harsh environments. Table \ref{th} shows applications where thermal energy is utilized for powering up different devices. It is apparent from Table \ref{th} that a harvesting power in the range of hundreds of milliwatts is possible using thermal sources and could be potentially used for various applications.
\begin{table}[]\scriptsize
	\centering
	\caption{Thermal energy harvesting applications}\centering
	\label{th}
	\begin{tabular}{clcc}
		\toprule
		\textbf{Applications }                  & \multicolumn{1}{c|}{\textbf{Heat utilized}}              & \textbf{Harvested Energy} & Ref.\\ \midrule
		Seiko Thermic watch    & Wrist and the environment at room   temperature & $22$ µW   &   \citep{XinLu2010}    \\ 
		Nuclear   Power Plant   & Heat pipes                   & -   &\citep{Chen2016a}           \\ 
		ThermaWatt             & The heat of candle and room temperature         & $500-800$ mW&\citep{therm}      \\ 
		DW-DF-10W Camp Stove  & The heat of Propane stove                       & -   &\citep{DW}           \\ 
		Radiator     & Radiator $323$ $^{\circ}$K \& Air Voltage Current   Power $294$ $^{\circ}$K                                                   & $95.19$ mW &\citep{Franciscatto2014}\\ 
		Pavement     & \begin{tabular}[c]{@{}l@{}}The temperature difference between the pavement   \\ surface and the subgrade soil\end{tabular} & $0.05$ mW &\citep{Wu2012} \\ 
		Aircraft                & Cargo   skin and Cargo primary insulation       & $22.58$ mW  &  \citep{Pearson2012}    \\
		 \bottomrule 
	\end{tabular}
\end{table}

The thermoelectric generators are generally manufactured from either inorganic or polymer materials.   The inorganic materials are mostly based on Bi-Te compounds. The thermoelectric generator consists of inorganic bulk materials embedded in a flexible polymer. The flexible thermoelectric generator can be attached to a curved surface as well. Table \ref{eff} shows the conversion efficiency for different thermoelectric generator materials. It is well known that the conversion efficiency of thermoelectric generators is very low, making them unsuitable for various standalone applications.  

The performance of thermoelectric material is usually measured using a dimensionless figure of merit known as ZT value. The ZT value is directly proportional to the Seebeck coefficient and electrical conductivity. In order to improve the conversion efficiency of thermoelectric generators, a high value of ZT at room temperature is desired. The maximum conversion efficiency of $84.5\%$ is achieved when the ZT of the material is $1.8$ at $298$ $^o$K as indicated in Table \ref{eff}.
\begin{table}[]\scriptsize
	\caption{Conversion efficiency for different thermoelectric generator compositions}
	\label{eff}
	\centering
	\begin{tabular}{ccccc}
		\toprule
		\textbf{	Material }                       &\textbf{ ZT (at $298$ $^{\circ}$K) }& \textbf{$\Delta$T} & \textbf{Conversion efficiency ($\%$)} &\textbf{ Ref.}     \\ \midrule
		Bi$_2$Te$_3$                    & $0.69$                    & $137$     & $54.6$                       & \citep{Chen2015a} \\ 
		Bi$_2$Te$_3$, Sb$_2$Te$_3$       & -                         & $15$      & $43$                         & \citep{Jang2011}  \\ 
		(BiSb)$_2$Te$_3$, Sb$_2$Te$_3$ & -                         & $240 $    & $81.8$                       & \citep{Barma2015}\\ 
		Bi$_2$(Te,Se)$_3$               & $1$                       & $125$     & $60.4$                       & \citep{Chen2016b} \\ 
		(Bi,Sb)$_2$ Te$_3$              & $1.4$                     & $125$     & $74$                         & \citep{Chen2016b} \\ 
		Bi$_2$Te$_3$   super-lattices   & $1.8$                     & $125$     & $84.5 $                      & \citep{Chen2016b} \\ \bottomrule
	\end{tabular}
\end{table}

\begin{figure}[h]
	\centering
	\includegraphics[width=7.2cm]{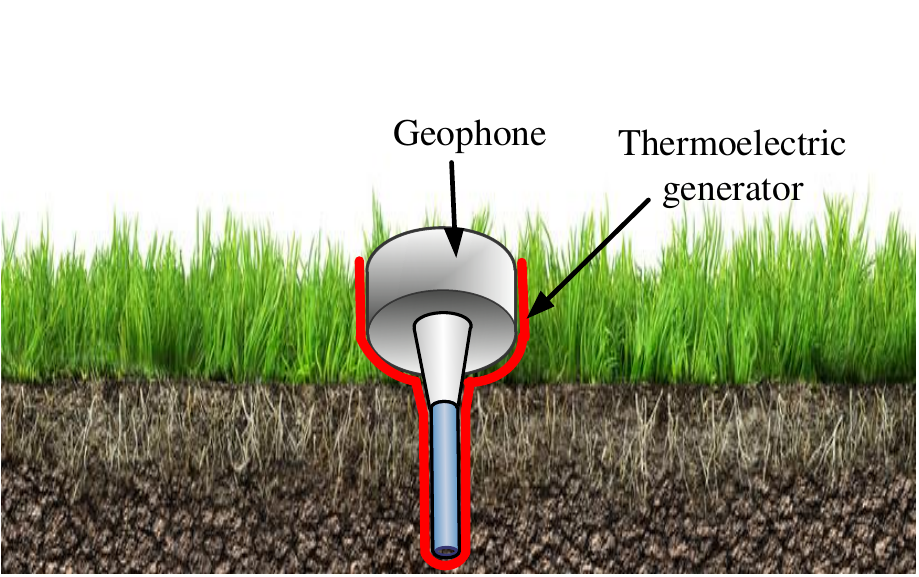}\\
	\caption{Geophone with the thermoelectric generator.}
	\label{fig:geo_TEG}
\end{figure}

Thermoelectric generators generally require a temperature gradient of around $5-10$ $^{\circ}$K to generate electrical power in the milliwatt range \citep{Chen2015a}. Here, we propose to use thermoelectric generators to be placed on the outer surface of the geophones installed in the seismic field. As shown in Fig. \ref{fig:geo_TEG}, a part of the geophone is under the ground. This creates a temperature gradient due to the temperature difference between the ground and the subsurface. Usually, a significant temperature difference exists between the upper surface of the seismic field and few centimeters below it. The harvested energy from the thermoelectric generator can be utilized to provide power to geophones installed in seismic fields.   A closely related scenario has been recently studied by Sigrist et.al. \citep{Sigrist2020}. The authors in the aforementioned work developed an end-to-end thermoelectric energy harvesting system to harvest energy from temperature gradients found at the natural ground-to-air boundary on the earth's surface. Table \ref{gta} lists various thermoelectric generators  that generate power using ground-to-air temperature gradient.
\subsection{Discussion}
In order to demonstrate the feasibility of thermal energy harvesters in seismic fields, we have recorded the temperature during the month of November in Dammam, Saudi Arabia (see Table \ref{temp_diff}) and found out the temperature difference of about $5 - 7$ $^{\circ}$K exists $10$ cm below the surface.  We strongly believe that this temperature difference will generate significant power to contribute to the energy required by geophones. This  energy harvesting source is readily available for $24$ hours a day and can account for the significant portion of energy harvested from various sources. Moreover, the thermal energy harvesters are also robust to high temperatures and dust. 

\begin{table}[]\scriptsize
	\caption{Thermoelectric generators utilizing ground-to-air temperature gradient}
	\label{gta}
	\begin{tabular}{cccc}
		\toprule
		\textbf{\begin{tabular}[c]{@{}c@{}}Thermoelectric \\ Generator characteristics\end{tabular}} &
		\textbf{\begin{tabular}[c]{@{}c@{}}Power \\ output (mW)\end{tabular}} &
		\textbf{Power type} &
		\textbf{Ref.} \\ \midrule
		Area of $144$ cm$^2$ &
		$1.1$ &
		Average &
		\citep{Whalen2012} \\
		\begin{tabular}[c]{@{}c@{}}Optimized source and load\\ matching\end{tabular} &
		$8.1 \times 10^{-4}$ &
		Average &
		\citep{Moser2012} \\
		\begin{tabular}[c]{@{}c@{}}Finned thermal guides of $3.8$ cm\\ diameter\end{tabular} &
		$1$ &
		Average &
		\citep{Stevens2013} \\
		Area of $80$ cm$^2$ &
		$16$ &
		Average &
		\citep{Datta2018} \\
		Only Simulations  setup &
		\begin{tabular}[c]{@{}c@{}}Enough to power \\ a wireless sensor\end{tabular} &
		- &
		\citep{Pullwitt2018} \\
		\begin{tabular}[c]{@{}c@{}}With active rectification circuit \\ and electrical impedance matching\end{tabular} &
		\begin{tabular}[c]{@{}c@{}}$27.2$  (day) $6.3$ (night)\\ 1.1\end{tabular} &
			\begin{tabular}[c]{@{}c@{}}Peak \\ Average \end{tabular} &
		\citep{Sigrist2020a}\\ \bottomrule
 	\end{tabular}
\end{table}
\begin{table}[]\scriptsize
	\caption{Temperature measurements in the Eastern region of Saudi Arabia during the month of November.}
	\label{temp_diff}
	\begin{tabular}{ccccc}
		\toprule
		\textbf{Time} &
		\begin{tabular}[c]{@{}c@{}}\textbf{Ambient }\\ \textbf{temperature}\end{tabular} &
		\begin{tabular}[c]{@{}c@{}}\textbf{Temperature} \\ $1$ \textbf{cm below} \\ \textbf{surface}\end{tabular} &
		\begin{tabular}[c]{@{}c@{}}\textbf{Temperature} \\ $5$ \textbf{cm below} \\ \textbf{surface}\end{tabular} &
		\begin{tabular}[c]{@{}c@{}}\textbf{Temperature} \\ $10$ \textbf{cm below} \\ \textbf{surface}\end{tabular} \\ \midrule
		Morning & $293$ $^{\circ}$K & $289$ $^{\circ}$K & $289$ $^{\circ}$K & $288$ $^{\circ}$K \\ 
		Noon    & $306$ $^{\circ}$K & $313$ $^{\circ}$K & $307$ $^{\circ}$K & $299$ $^{\circ}$K \\ 
		Evening & $293$ $^{\circ}$K & $294$ $^{\circ}$K & $298$ $^{\circ}$K & $300$ $^{\circ}$K \\ 
		Night   & $293$ $^{\circ}$K & $290$ $^{\circ}$K & $294$ $^{\circ}$K & $298$ $^{\circ}$K \\ \bottomrule
	\end{tabular}
\end{table}

\section{RF Energy Harvesting}\label{rfh}
\begin{figure}[h]
	\centering
	\includegraphics[width=7.2cm]{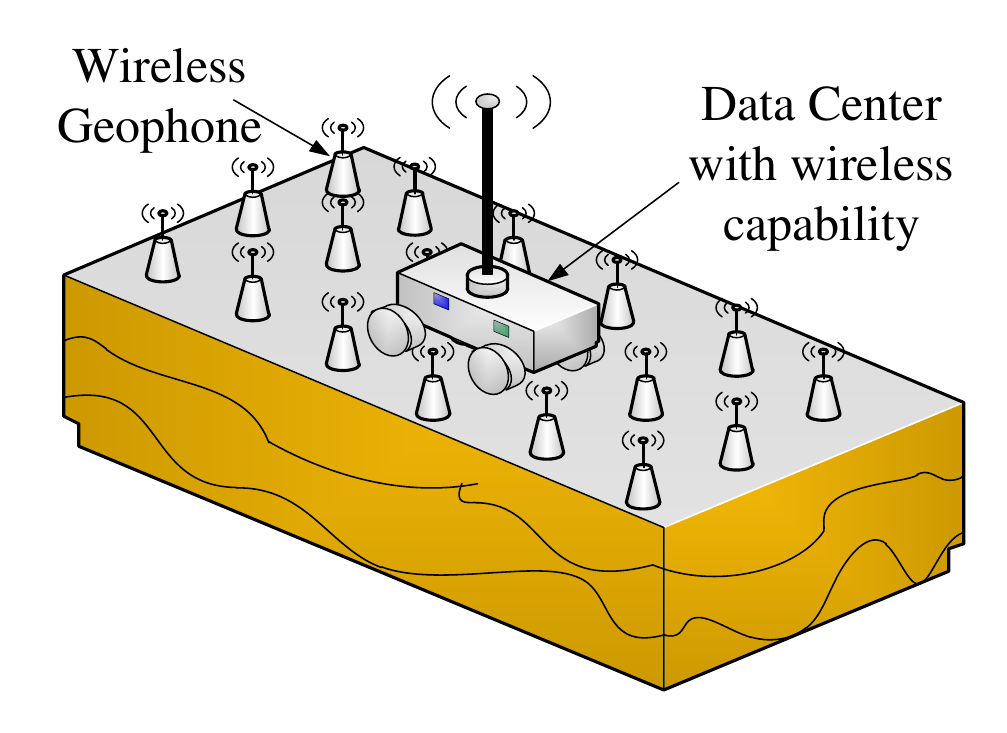}\\
	\caption{Geophones and data center in the seismic field.}
	\label{fig:geo_layout}
\end{figure}

 Harvesting energy from RF sources, also known as wireless energy harvesting,  has found a lot of interest in recent years due to their wide applications as a substitute power source for various applications. Some interesting applications include battery-less power sources, RF tags, biomedical devices, and smart wireless sensor networks which require nanowatt to microwatt input power. We believe that wireless geophones can also take advantage of this technology. Specifically, the presence of an on-site data center provides an opportunity to power wireless geophones through RF energy. A typical layout of geophones in the field and an on-site data center is shown in Fig. \ref{fig:geo_layout}.  Power is readily available at the data centers and can be used to transmit energy to geophones using a wireless link.

\subsection{RF Energy Sources}
In general, a wireless geophone can harvest RF energy from various different sources. Any device emitting radio waves can be considered as a source for wireless energy harvesting. The frequency range of such sources depends on the type of transmitter. The most common radio sources are radio/TV broadcasting stations, satellites, wireless fidelity (Wi-Fi), global system for mobile communications (GSM), universal mobile telecommunications system (UMTS), and long term evolution (LTE) base stations. These sources cover a broad range of frequencies starting from $3$ kHz to all the way up to $300$ GHz of the electromagnetic spectrum. These RF energy sources are ubiquitous and are even available in the most inaccessible places. 

A typical RF energy harvesting system consists of an antenna that receives the incident power, a matching network for maximizing the power transfer and minimizing the signal reflection, and an RF to DC rectifier  \citep{ParejaAparicio2016}. RF energy harvesting can also be used along with data transfer in a communication system.  Table \ref{pd} shows the power density of different RF sources. The power densities of these sources vary from $0.45$ nW/cm$^2$ for GSM900 mobile terminal to $84$ nW/cm$^2$ for GSM1800 base station. 

\begin{table}[]\scriptsize
	\centering
	\caption{Power density of different RF sources (DTV: Digital TV; MT: Mobile Terminal; BT: Base Terminal)}
	\label{pd}
	
	\begin{tabular}{ccc}
		\toprule
		\textbf{Band}         & \textbf{Range (MHz)}    & \textbf{Power Density (nW/cm$^2$)} \\ \midrule
		DTV          & $470-610 $  & $0.89$                 \\ 
		GSM900 (MT)  & $880-915$   & $0.45 $                 \\ 
		GSM900 (BT)  & $920-960$   & $36$                   \\ 
		GSM1800 (MT) & $1710-1785$ & $0.5 $                  \\ 
		GSM1800 (BT) & $1805-1880$ & $84$                   \\ 
		3G(MT)       & $1710-1785$ & $0.46$                    \\ 
		3G(BT)       & $2110-2170$ & $12 $                   \\ 
		WiFi         & $2400-2500$ & $0.18$                  \\ \bottomrule
	\end{tabular}
\end{table}

\begin{table}[]\scriptsize
	\caption{Conversion efficiency for different RF schemes}
	\label{rf}
	\centering
	\begin{tabular}{ccccccc}
		\toprule
		\begin{tabular}[c]{@{}c@{}}\textbf{Frequency}  \\ \textbf{Band} (MHz)\end{tabular} &
		\begin{tabular}[c]{@{}c@{}}\textbf{Type of}  \\ \textbf{antenna} used\end{tabular} &
		\begin{tabular}[c]{@{}c@{}}\textbf{Input } \\ \textbf{Power (dBm)}\end{tabular} &
		\begin{tabular}[c]{@{}c@{}}\textbf{Output } \\ \textbf{Voltage (V)}\end{tabular} &
		Load ($\omega$) &
		\begin{tabular}[c]{@{}c@{}}\textbf{Conversion}  \\ \textbf{efficiency ($\%$)}\end{tabular} &
		\textbf{Ref.} \\ \midrule
		$0.47-0.86$        & \begin{tabular}[c]{@{}c@{}}Just rectifier,\\  no antenna used\end{tabular}   & $10   $  & $-   $   & $12200   $ & $60   $ & \citep{Bolos2016}\\ 
		$0.9 - 2.45   $  & Patch                                                                         & $-3   $  & $21   $  & $2400   $  & $50   $ & \citep{Marian2012} \\ 
		$0.876 - 0.959   $ & Dipole                                                                        & $5.8   $ & $0.9   $ & $11000   $ & $84   $ &\citep{Kuhn2015} \\ 
		$0.9   $           & Patch                                                                         & $-15   $ & $-   $   & $50000   $ & $45   $ & \citep{Keyrouz2013} \\ 
		$1.5 - 2.0   $     & \begin{tabular}[c]{@{}c@{}}Just rectifier,\\  no antenna used\end{tabular}    & $27   $  & $-   $   & $50   $    & $55   $ & \citep{Oka2014} \\ 
		$0.85   $          & Patch                                                                         & $-20   $ & $-   $   & $2200   $  & $15   $ & \citep{Collado2013} \\ 
		$2.45   $          & Microstrip                                                                    & $0   $   & $1   $   & $1400   $  & $83   $ & \citep{HuchengSun2012} \\ \bottomrule
	\end{tabular}
\end{table}

Table \ref{rf} shows the reported conversion efficiencies of different RF schemes with various antenna types. In \citep{Bolos2016} an RF energy harvester is reported where there is no antenna and a uniform transmission line was used for impedance matching. It is evident that dipole and microstrip antenna offered the best conversion efficiency followed by schemes based on patch antennas.

\subsection{Optimal Signal Design for RF energy harvesting}\label{otsi}
The signal waveform design also plays an important role in efficient RF energy harvesting. Various waveform designs based on single or multiple antenna transmissions are reported in the literature \citep{Clerckx2016,Kim2019}. It has been shown that the design of an appropriate signal generation method that adapts as a function of the channel condition, significantly boosts the amount of harvested energy \citep{Kim2019}. Particularly, the transmitted RF signal has been proposed to be the superposition of multiple sine-waves of unique amplitudes and phases, where the number of sine-waves depends upon the number of channel subbands.

Consider a general multiple-antenna transmitter with $M$ transmit antennas and assume $N$ channel subbands for a general frequency-selective channel. The transmit vector signal can be expressed as \citep{Kim2019}
\begin{align}
\mathbf{x}(t)&=\Re\left\{\sum_{n=0}^{N-1} \mathbf{w}_{n} e^{j 2 \pi f_{n} t}\right\}
\end{align}
where $\mathbf{x}(t)=\left[x_{1}(t), \cdots, x_{M}(t)\right]^{T} \quad$ is a vector of transmitted signal from $M$ antennas, $\mathbf{w}_{n}=$
$\left[w_{n, 1}(t), \cdots, w_{n, M}(t)\right]^{T}$ with $w_{n, m}(t)=s_{n, m}(t) e^{j \phi_{n, m}(t)}$ expresses the amplitude and phase of the subband
signal on frequency $f_{n}$ and transmit antenna $m$ at time $t$. If the frequency response of the multipath channel
is given by $h_{n, m}=A_{n, m} e^{j \psi_{n, m}}$,  the optimal design of  $\mathbf{w}_{n}$  is given by \citep{Kim2019}
\begin{align}\label{d1}
\mathbf{w}_{n}=\frac{\mathbf{h}_{n}^{H}}{\left\|\mathbf{h}_{n}\right\|}\left\|h_{n}\right\|^{\beta} \sqrt{\frac{2 P}{\sum_{n=0}^{N-1}\left\|\mathbf{h}_{n}\right\|^{2} \beta}}
\end{align}
where $\mathbf{h}_{n}=\left[h_{n, 1}, \cdots, h_{n, M}\right]$, and $\beta$ is a scaling factor whose optimal value is chosen to be $3$ \citep{Kim2019}, and $P$ is the
transmit power budget. Under a single-antenna transmitter, the optimal design can be expressed as
\begin{align}\label{d2}
w_{n}=A_{n}^{\beta} \sqrt{\frac{2 P}{\sum_{n=0}^{N-1} A_{n}^{2}}} e^{-j \psi_{n}}
\end{align}

\subsection{Discussion}
In the seismic exploration environment, we propose that wireless geophones tap the RF energy generated by the data center. In the seismic acquisition, geophones transmit the acquired data to the data center, which in turn sends acknowledgments \citep{Iqbala} usually in the form of small frames. The fact that only small frames are sent in the downlink, makes it a perfect scenario to harvest energy from the RF signals. Since  the downlink channel is idle most of the time, special signals (as discussed in Section \ref{otsi}) meant for energy harvesting can be sent over it. Moreover, geophones can be powered up using RF signals during both the shooting interval and the non-shooting periods. It means that RF signals can be used to power up geophones at any time. 

In \citep{Iqbala}, the authors proposed a scheme for seismic data transmission utilizing wireless network based on IEEE802.11af standard. Usually the ambient energy from this RF source is not sufficient for powering the geophonesss and, therefore, other sources need to be added to the system. Nevertheless, we still believe that it can be utilized with other energy harvesting modes in a hybrid fashion. Furthermore, in the wireless sensor network literature, there is a growing trend of using unmanned aerial vehicles (UAV) to power up sensor nodes through RF signals \citep{8422707, 9027446, 8742610}. The same concept can be applied to power up geophones located far away from the data center where RF energy harvesting is not feasible. In such large distance scenarios as geophone cannot transmit the recorded data to data centers directly, UAVs are sent to collect the data \citep{Stephenson2019}. Thus the UAVs can be used to simultaneously receive data from and transmit power to the geophones.

As mentioned above that downlink contains only acknowledgments. The \emph{almost} idle channel can be leveraged to intelligently design waveforms that are friendly for RF energy harvesting operation. Thus the amount of energy being harvested can be improved for the geophones. Another interesting design strategy could be to use these special waveforms that can maximize the RF energy harvesting efficiency  as acknowledgments (positive or negative) for a geophone. Finally, the waveform design in (\ref{d2}) suggests multiple antennas at the data center and a single antenna at a geophone. This is perfect for a typical wireless seismic acquisition setup since it relieves a limited-power geophone while shifting heavy processing to the  data center where power requirements are relaxed.

\section{Proposed Design of Energy Harvesting Geophone}\label{prop}
\begin{figure}[h]
	\centering
	\includegraphics[width=12.2cm]{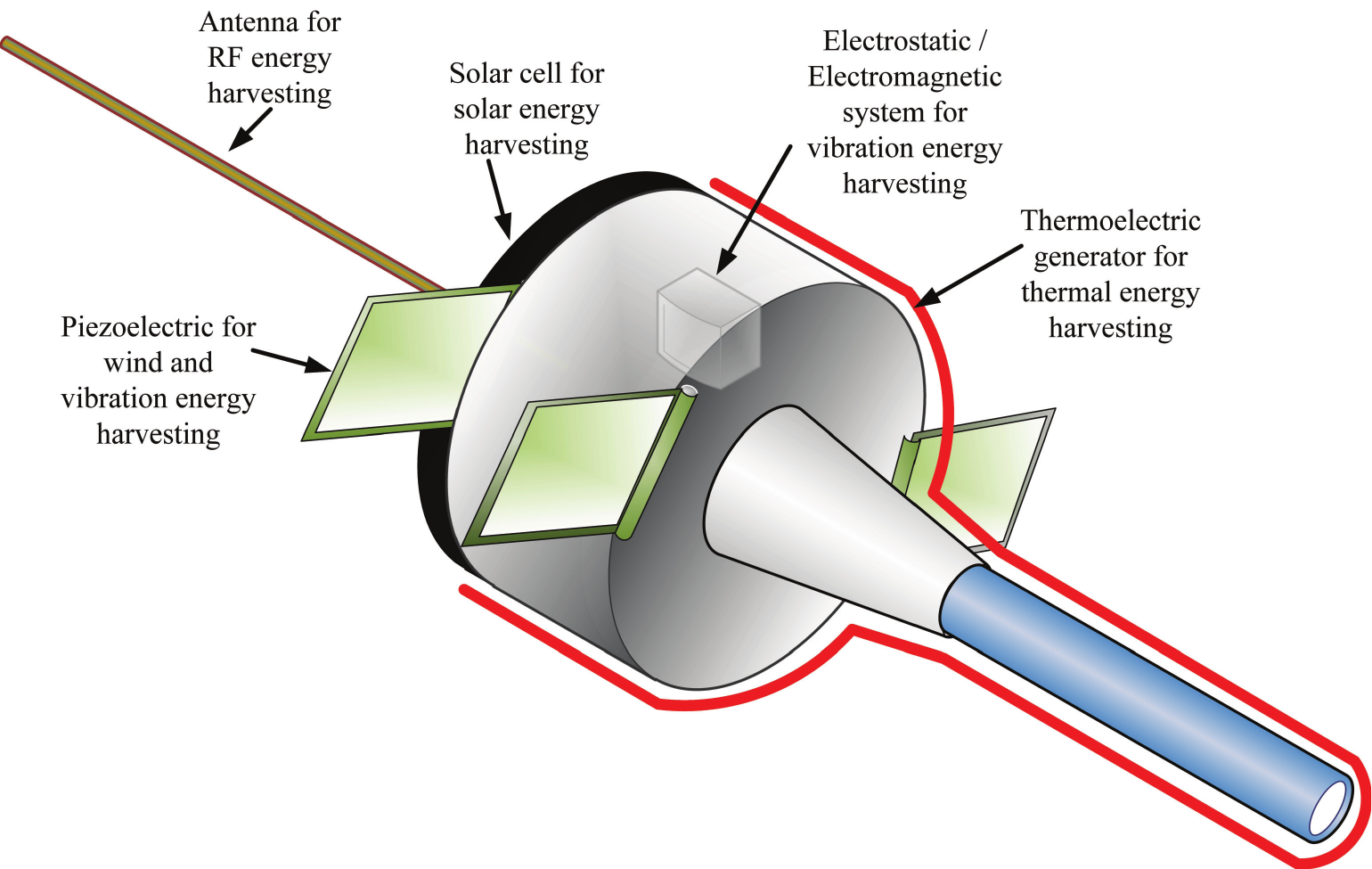}\\
	\caption{Cross-section of a geophone with various energy harvesting systems.}
	\label{fig:geophone_eh}
\end{figure}

Figure \ref{fig:geophone_eh} illustrates the proposed design of an energy harvesting geophone. It consists of solar cells on the top surface, piezoelectric and electromagnetic/electrostatic harvesters on the sides/edges and inside, respectively, coating of thermoelectric material on the whole body and antenna for RF energy harvesting.

Based on this design, the average energy harvested per day can be approximated using the appropriate harvester designs highlighted in the above Sections \ref{so}-\ref{rfh}. We have selected a suitable harvester per energy source based on the latest design, practicality, size, feasibility to geophones, maximum efficiency, and output power.
 From the Table \ref{all}, it can be concluded that the average harvested energy from multi-source can be used to meet the power requirements of a geophone. 
 Furthermore, RF energy harvester is the least effective in this case, while solar energy harvesters contribute to most of the harvested energy. Although RF energy harvester adds very little to the overall harvested energy, we strongly believe that it can still be useful as geophones can be powered up by RF energy (transmitted from the data center) for $24$ hours a day  especially during the evening/night when geophones are in sleep mode (no recording). In addition, recently, there are considerable improvements in the conversion efficiency of RF circuits (e.g. see \cite{Cansiz2019,Assogba2020}) which could improve their performance in the future. 
The proposed design is a collective solution to harvest energy through various possible means. One can easily adapt to pick and choose some and drop others, depending upon the actual environmental situation. The hardware implementation and extensive comparison of various harvester designs in the real scenario is the focus of our future research.

\begin{table}[]\scriptsize
\caption{Energy harvesting by a geophone during $24$ hours in seismic field}
\label{all}
\begin{tabular}{cclcc}
\toprule
Energy source &
  Harvester design &
  Description &
  \begin{tabular}[c]{@{}c@{}}Time duration of \\ energy availability\end{tabular} &
  \begin{tabular}[c]{@{}c@{}}Average energy \\ harvested\end{tabular} \\ \midrule
  Solar &
  \citep{Luceno-Sanchez2019}  &
  \begin{tabular}[l]{@{}l@{}}Duration of sunlight is taken from $10$ am to $2$ pm and area \\ is assumed to be $50$ cm$^2$ which is suitable for a geophone\end{tabular} &
  4 hours &
  $249$ J \\
  Vibration &
  PPA-4011 &
  \begin{tabular}[l]{@{}l@{}}Acquisition is done from $8$ am to $4$ pm and shots are  \\  carried out in interval of $8$ sec, hence vibration \\ energy is available  during the shots\end{tabular} &
  \begin{tabular}[l]{@{}l@{}}1 hour and \\ $75$ mins\end{tabular} &
  $51$ J \\
  RF &
  WiFi Band &
  \begin{tabular}[l]{@{}l@{}}Geophones can be powered up by RF for the whole $24$ \\ hours  and the antenna is  assumed to be $1$ m$^2$\end{tabular} &
  24 hours &
  $0.15$ J \\
  Thermal &
  \citep{Sigrist2020a} &
  Available $24$ hours &
  24 hours &
  $95$ J \\
  Wind &
  \citep{Park2014} &
  Available $24$ hours with average speed of $\approx 4$ m/s &
  24 hours &
  $64.37$ J \\ \bottomrule
\end{tabular}
\end{table}
\section{Conclusion}\label{conc}
This paper has presented a comprehensive survey of the promising energy harvesting technologies for realizing self-powered geophones for seismic exploration. First,  an overview of a typical wireless geophone with a focus on its energy requirements is provided. Next, detailed discussions about the state-of-the-art research contributions in various small-scale energy harvesting techniques suitable for geophones are presented. These included solar, vibration, wind, thermal, and RF energy harvesting methods. To this end, characteristics and design of different energy harvesting methods, their limitations, amount of harvested energy, comparisons, and various research challenges are discussed along with some real case studies. Finally, we have outlined the proposed design of a geophone equipped with the discussed energy harvesting mechanisms. It is concluded that energy harvesting and storage systems are to be planned based on the combination of more than one alternative energy source. It paves the way for a paradigm shift from traditional wired geophones to a truly autonomous and sustainable geophone energy harvesting network. The hardware design and implementation issues were identified as future research directions. We believe these insights will motivate further research towards the use of energy harvesting in geophones.

	\bibliographystyle{ieeetr}
	\bibliography{library_fixed,SolarBibliography,SeismicHarvesting_fixed}

\newpage

\end{document}